\documentclass[aip,jcp,showpacs,preprintnumbers,amsmath,amssymb,floatfix,a4paper,reprint,longbibliography,superscriptaddress,citeautoscript]{revtex4-2}

\usepackage{amsmath}
\usepackage{amsfonts}
\usepackage{amssymb}
\usepackage{bm}
\usepackage{bbm}
\usepackage{color}

\usepackage{graphicx}

 \usepackage{hyperref}
\hypersetup{breaklinks=true}
\usepackage{dcolumn}
\usepackage{array}
\usepackage{diagbox}
\usepackage{orcidlink}

\usepackage[T1]{fontenc}   %Either of these two packages are needed to get the Polish ogonek (a small hook beneath an a)
\newcommand{\figwidth}{1\columnwidth}%for two-column
\newcommand{\figwidthbroad}{2\columnwidth}%broad figure spanning two columns
\newcommand{\nqubit}{N_{\rm q}}
\newcommand{\nqubitrot}{N_{\rm q}^{\rm rot}}
\newcommand{\nqubitvib}{N_{\rm q}^{\rm vib}}
\newcommand{\vmax}{v_{\rm max}}
\newcommand{\Ntrotter}{N_{\rm ST}}
\newcommand{\Natoms}{N_{\rm a}}
\newcommand{\Nmodes}{N_{\rm vib}}
\newcommand{\Nshots}{N_{\text{shot}}}
\newcommand{\NHq}{L_{\text{q}}}
\newcommand{\Hsize}{S}
\newcommand{\UST}{U}

\DeclareMathOperator{\bin}{bin}
\DeclareMathOperator{\trace}{tr}
\newcommand{\bigO}{\mathcal{O}}
\newcommand{\Reimei}{Reimei}
\newcommand{\vmaxexponent}{\eta}
\newcommand{\hcmminusone}{$hc\,$cm$^{-1}$}

\begin{document}

\title{Rovibrational energy levels of H$_2$O by quantum computing}
\date{\today} 
%\preprint{RIKEN-iTHEMS-Report-26}
\author{Erik L\"otstedt\,\orcidlink{0000-0003-2555-8660}}
%\author{Erik L\"otstedt}
\email{loetstedte@riken.jp}
\affiliation{RIKEN Center for Interdisciplinary Theoretical and Mathematical Sciences (iTHEMS), Wako, Saitama 351-0198, Japan}
\affiliation{Trapped Ion Quantum Computer Team, TRIP Headquarters, RIKEN,  Wako, Saitama 351-0198, Japan}
\affiliation{Computational Condensed Matter Physics Laboratory, RIKEN Pioneering Research Institute (PRI), Wako, Saitama 351-0198, Japan}
 \author{Tam\'as Szidarovszky\,\orcidlink{0000-0003-0878-5212}}
 %\author{Tam\'as Szidarovszky}
 \email{tamas.janos.szidarovszky@ttk.elte.hu}
\affiliation{Institute of Chemistry, ELTE E\"otv\"os Lor\'and University, P\'azm\'any P\'eter S\'et\'any 1/A, H-1117 Budapest, Hungary}

\begin{abstract}
We calculate rovibrational energy levels of H$_2$O using a trapped-ion quantum computer. We first derive the qubit form of Watson's  Hamiltonian, including the rovibrational coupling terms. In a second step, we employ a variant of the quantum-selected configuration-interaction method to calculate rovibrational energy levels. A truncated form of the qubit Hamiltonian is used to generate correlated rovibrational wave functions on the quantum computer by time evolution, and a basis set is selected  by sampling from the measured  probability distribution.  The rovibrational energy levels are obtained by constructing a Hamiltonian matrix using the selected basis set, and diagonalizing the matrix using a classical computer. 
We show that  an accuracy of a few \hcmminusone{} can be achieved for low-lying rovibrational energy levels.
\end{abstract}

%\keywords{}

\maketitle
%
%\tableofcontents
%\newpage %Only needed for preprint style

%=========================================================================================================================
\section{Introduction}
%=========================================================================================================================
The electronic structure problem, that is, the calculation of wave functions and energies of the electrons in a molecule, has so far been the main focus of research on quantum computing applied to chemistry
\cite{Caoetal_review2019,Baueretal2020,McArdleetal_review2020,Claudino2022,Pyrkov2023,Mazzola2024,%
Weidman2024,Patel2025, Alexeev2025, Schleich2025,Zhang2025,Alfonso2025}. Although it is believed that the solution to the electronic structure problem will be one of the main applications for future fault-tolerant quantum computers  \cite{Katabarwa2024}, it is difficult to calculate electronic energies of large molecules using currently available quantum computers due to inherent noise. Thus far, only small molecules such as Li$_2$O \citep{ZhaoGoings2023}, F$_2$ \cite{Guo2024}, and SrH \cite{Chawla2025} have been simulated.
Recently, larger molecules such as  [Fe$_4$S$_4$(SCH$_3$)$_4$]$^{2-}$  \cite{RobledoMoreno2025,Shirakawa2025} and C$_6$H$_{12}$ \cite{Shajan2025} have been treated with hybrid methods in which the quantum computer is used to select the active space, while  a classical computer is used for the energy evaluation.

On the other hand, to calculate molecular energy levels, we must  solve also the  Schr\"odinger equation for the coupled vibrational and rotational motion of the molecule \cite{Csaszaretal2012}.  The solution of the rovibrational Schr\"odinger equation is  a difficult problem: the number of  variables $N_{\text{var}}$ on which the rovibrational wavefunction depends increases linearly with the number of atoms $\Natoms$ as $N_{\text{var}}=3\Natoms -3$, corresponding to $3\Natoms -6$ vibrational and three rotational coordinates. 
By using a quantum computer, the hope is that by mapping the rovibrational wave function to qubit form which can be represented on a quantum computer, we can obtain accurate energy levels also for molecules too large to simulate using classical computers.
However, compared to the large body of  literature on the application of quantum computers to the electronic structure problem, there are relatively few reports on the application of quantum computers to the rovibrational structure problem. Pioneering studies on the vibrational structure of small molecules such as O$_3$ \cite{Teplukhin2019,Sawaya2021} 
H$_2$O \cite{McArdleetal2019}, CO$_2$ \cite{Ollitraultetal2020,Loetstedt2021,Loetstedt2022_copy,Nguyen2023,Somasundarametal2025}, Cr$_2$ \cite{Asnaashari2026} and NH$_3$ \cite{Nguyen2023,Somasundarametal2025} have been carried out. There are also proposals for evaluating the Franck-Condon factors in vibronic transitions  \cite{Huhetal2015,Shen2018,SawayaHuh_2019,Jahangirietal_2020,Wangetal2020}, as well as investigations on the simulation of time-dependent vibrational wave packets 
\cite{Sparrow2018,Lee2022,Malpathak2025}.

In experimental spectroscopy, transitions among rovibrational energy levels   of small molecules in the gas phase  can be measured with very high precision \cite{HandbookHighresSpectroscopy2011,Shumakova2024,SilvadeOliveira2024}. The uncertainty of the wave number of a measured transition is often a fraction of a cm$^{-1}$, and in many cases as small as $10^{-6}$ cm$^{-1}$. 
In energy units, we have 1 \hcmminusone{} $\approx 0.12$ meV $\approx 4.6\times 10^{-6}E_{\rm h}$, where $h$ is  Planck's constant, $c$ is the speed of light,  and $E_{\rm h}\approx 27.211$ eV is 1 hartree. 
``Spectroscopic accuracy'' in a theoretical calculation of a rovibrational energy level is commonly referred to as an accuracy of 
1 \hcmminusone{} \cite{Peterson2012,Fortenberry2016}.
The experimental transition frequencies are collected in databases such as HITRAN \cite{HITRAN2022,Gordon2026}, and in the case of well-studied molecules such as H$_2$O \cite{Furtenbacher2020a,Furtenbacher2020} or NH$_3$ \cite{Furtenbacher2020b}, there are  thousands of measured transitions.
To interpret the experimental data we must calculate the rovibrational energy levels involved in the transitions, which is a difficult task for a polyatomic molecule. To obtain a good agreement with the experimental data, it is essential to include the coupling between the rotational and vibrational degrees of freedom. However, 
 in  previous  investigations on vibrational structure evaluation by quantum computing \cite{Teplukhin2019,Sawaya2021,McArdleetal2019,Ollitraultetal2020,Loetstedt2021,Loetstedt2022_copy,%
 Nguyen2023,Asnaashari2026,Nguyen2023,Somasundarametal2025,Majland2026},
  the rotational motion of the molecule was not considered. Very recently, Szczepanik and coworkers presented an algorithm for the evaluation of rovibrational wave functions using fault-tolerant quantum computers \cite{Szczepanik2025}.

In this paper, we aim to calculate rovibrational energy levels using currently available noisy quantum computers, making our investigation complementary to that for fault-tolerant quantum computers presented in Ref.~\onlinecite{Szczepanik2025}.
We derive the qubit form of the rovibrational Hamiltonian and discuss the relative importance of the different terms in the Hamiltonian. Taking the water molecule as an example, we employ the classical-quantum hybrid 
quantum-selected configuration interaction (QSCI) method \cite{Kanno2026,Nakagawa2024} and evaluate a number of rovibrational energy levels using the Quantinuum trapped-ion quantum computer \Reimei{} \cite{QuantinuumReimei2025}. 
In standard methods for eigenvalue evaluation such as the variational quantum eigensolver (VQE) method\cite{Peruzzoetal2014,Yuan2019,Lietal2019,Parrishetal2019,Cerezo2021}, where the expectation value of the Hamiltonian is evaluated using a quantum computer, and the quantum subspace expansion (QSE) method
\cite{McCleanetal2017,Yoshioka2022,OLeary2025},  where the matrix elements of the Hamiltonian are evaluated using a quantum computer, the resulting energies are sensitive to noise and require  an excessive number of measurements.
On the other hand, in the QSCI method, the Hamiltonian matrix is constructed using a classical computer. This makes the QSCI method applicable to  
 problems in chemistry that are difficult to treat with VQE-type methods using noisy quantum computers.

%=========================================================================================================================
\section{Theory}\label{Sec:Theory}
%=========================================================================================================================

%---------------------------------------------------------------------------------------------------------------------------------------------------------------------------------------------------------------
\subsection{Rovibrational Hamiltonian}\label{Subsec:RovibH}
%---------------------------------------------------------------------------------------------------------------------------------------------------------------------------------------------------------------
In this section, we review some parts of rovibrational structure theory.  An extensive review can be found in Ref.~\onlinecite{Bauder2011}.

We consider a nonlinear polyatomic molecule having $\Nmodes$ ($=3\Natoms -6$) vibrational normal modes, described by mass-scaled normal-mode coordinates $Q_1,Q_2,\ldots,Q_{\Nmodes}$ and conjugate momenta $P_k = -i\hbar\partial/\partial Q_k$. The rovibrational Hamiltonian,  derived by Darling and Dennison \cite{Darling1940} and simplified by Watson \cite{Watson1968}, is defined as
\begin{equation}\label{Eq:WatsonH}
H_{\rm W}= \frac{1}{2}(\bm{J}-\bm{\pi})^T\bm{\mu}(\bm{J}-\bm{\pi}) + \frac{1}{2}\sum_{k=1}^{\Nmodes}P_k^2 +V-\frac{\hbar^2}{8}\trace (\bm{\mu}).
\end{equation}
In Eq.~\eqref{Eq:WatsonH}, $\bm{J}=(J_a,J_b,J_c)^T$ is the angular momentum operator  in the rotating molecular frame, $T$ denotes the transpose of a vector, 
$\bm{\pi}=(\pi_a,\pi_b,\pi_c)^T$ is defined in terms of the Coriolis coupling coefficients  $\zeta_{kl}^\alpha$ ($\alpha=a$, $b$, $c$) as 
\begin{equation}
\pi_\alpha=\sum_{k,l=1}^{\Nmodes}\zeta_{kl}^\alpha Q_kP_l,
\end{equation}
$\bm{\mu}$ is the inverse inertia tensor, and $V$ is the potential energy surface. We provide detailed definitions and properties of $\bm{J}$, $\zeta_{kl}^\alpha$ and $\bm{\mu}$ in Appendix~\ref{Appendix:Defs}. The last term proportional to $\hbar^2$ in  the Hamiltonian \eqref{Eq:WatsonH} can be viewed as a quantum correction, because all other terms in \eqref{Eq:WatsonH} appear also in  the classical Hamiltonian for a rotating and vibrating molecule \cite{Bauder2011}.

The potential energy surface $V(Q_1,\ldots,Q_{\Nmodes})$ depends on all the normal mode coordinates in general. In the present investigation, we employ a  Taylor expansion up to fourth order in $Q_k$,
\begin{equation}\label{Eq:Vdef}
V(Q_1,\ldots,Q_{\Nmodes})=V_{\rm harm} + V_{\rm anharm},
\end{equation}
where 
\begin{equation}
V_{\rm harm} = \frac{1}{2}\sum_{k=1}^{\Nmodes} \omega_k^2 Q_k^2
\end{equation}
is the harmonic part of the potential defined in terms of the mode frequencies $\omega_k$, and
\begin{align}\label{Eq:VanharmDef}
V_{\rm anharm}={}&  
\frac{1}{6}\sum_{k,l,m=1}^{\Nmodes}g_{klm}Q_kQ_lQ_m \nonumber \\
&+\frac{1}{24}\sum_{j,k,l,m=1}^{\Nmodes}f_{jklm}Q_jQ_kQ_lQ_m
\end{align}
is the anharmonic part, defined in terms of
 the anharmonic coupling constants  $g_{klm}$ and $f_{jklm}$. We note that in the definition \eqref{Eq:VanharmDef}, there are many equivalent terms in the sum. For example, $g_{133}=g_{313}=g_{331}$ and so on.

In order to be consistent with the Taylor expansion of the potential energy, we also expand $\bm{\mu}$ up to fourth order in $Q_k$,
\begin{equation}\label{Eq:muexpansion}
\bm{\mu}=\sum_{\ell=0}^4\bm{\mu}_\ell,
\end{equation}
where  detailed expressions of the $\bm{\mu}_\ell$'s are given in Eqs.~\eqref{Eq:muorders}--\eqref{Eq:akDef} in Appendix.~\ref{Appendix:Defs}.
The actual Hamiltonian used in the simulations, $H$, is then defined by collecting all terms in $H_{\rm W}$ up to fourth order  in $Q_k$. To compare the relative importance of the respective terms, we group the terms in $H$ as follows,
\begin{equation}\label{Eq:Hdef}
H = H_{\rm RR}+ H_{\rm HO} + V_{\rm anharm} + H_{\rm vibCor} + H_{\rm  rovib},
\end{equation}
where
\begin{equation}\label{Eq:HRRDef}
H_{\rm RR} = \frac{1}{2}\bm{J}^T\bm{\mu}_0\bm{J}
\end{equation}
is the rigid-rotor Hamiltonian,
\begin{equation}
H_{\rm HO} =  \frac{1}{2}\sum_{k=1}^{\Nmodes}P_k^2 +V_{\rm harm} 
\end{equation}
is the harmonic-oscillator Hamiltonian, $V_{\rm anharm}$ is the anharmonic part of the potential energy surface defined in Eq.~\eqref{Eq:VanharmDef},
\begin{equation}\label{Eq:HvibCorDef}
H_{\rm vibCor}=\frac{1}{2}\sum_{\ell =0}^2\bm{\pi}^T \bm{\mu}_\ell \bm{\pi}  -\frac{\hbar^2}{8}\sum_{\ell = 0}^{4}\trace \bm{\mu}_\ell
\end{equation}
is the  part of the  Hamiltonian that describes the vibrational Coriolis coupling (in which we also include the small quantum correction term proportional to $\hbar^2$), and 
\begin{equation}\label{Eq:Hrovibdef}
H_{\rm  rovib} = \frac{1}{2} \sum_{\ell=1}^4 \bm{J}^T \bm{\mu}_\ell \bm{J} -\sum_{\ell =0}^3 \bm{J}^T \bm{\mu}_\ell \bm{\pi} 
\end{equation}
is the Hamiltonian describing the coupling of rotation and vibration.

Except for the recent Ref.~\onlinecite{Szczepanik2025}, in all previous investigations of quantum computing applied to molecular vibration 
\cite{Teplukhin2019,McArdleetal2019,Ollitraultetal2020,Sawaya2021,Loetstedt2021,Loetstedt2022_copy,%
Nguyen2023,Asnaashari2026,Nguyen2023,Somasundarametal2025}, only the $H_{\rm HO} +V_{\rm anharm}$ 
terms were considered. As we shall see in Sec.~\ref{Sec:Results}, including also the other terms is essential for obtaining accurate rovibrational energy levels.

%---------------------------------------------------------------------------------------------------------------------------------------------------------------------------------------------------------------
\subsection{Application to H$_2$O}
%---------------------------------------------------------------------------------------------------------------------------------------------------------------------------------------------------------------
The theory reviewed in the previous section \ref{Subsec:RovibH} applies to a general nonlinear polyatomic molecule. In the following, we apply the theory to the triatomic H$_2$O molecule as a concrete example. The water molecule has three vibrational modes: symmetric stretch ($\nu_1$), bending ($\nu_2$), and antisymmetric stretch ($\nu_3$). We evaluate the potential energy surface $V(Q_1,Q_2,Q_3)$
using  the CCSD(T) method \cite{Hampel1992,Deegan1994} and the aug-cc-pVQZ basis set \cite{Dunning1989,Kendall1992} implemented in {\sc Molpro} \cite{MOLPRO,Werner2011,Werner2020}.  The anharmonic coupling constants $g_{klm}$ and $f_{jklm}$ are evaluated using {\sc Molpro}'s  {\sc vpt2} program \cite{Ramakrishnan2015}. We give the numerical values of the relevant parameters in Appendix~\ref{Appendix:H2Oparameters}.

The rovibrational wave function is expanded as
\begin{equation}\label{Eq:Psiexpansion}
|\Psi\rangle =\sum_{v_1,v_2,v_3=0}^{\vmax}\sum_{K=-J}^J c_{v_1v_2v_3K}|v_1v_2v_3K\rangle,
\end{equation}
where the maximum vibrational quantum number $\vmax$ is assumed to be the same for all three modes, $c_{v_1v_2v_3K}$ is an expansion coefficient,  and  the basis state $|v_1v_2v_3K\rangle$ is defined as a direct product of harmonic oscillator eigenstates $|v\rangle$  and an angular momentum state $|J,K\rangle$, 
\begin{equation}
|v_1v_2v_3K\rangle = |v_1\rangle |v_2\rangle |v_3\rangle |J,K\rangle,
\end{equation}
where $K$ measures the projection of the total angular momentum onto the molecular $c$ axis. The quantum number $M$ of the projection of the total angular momentum on the lab-frame $z$ axis is not written out because the Hamiltonian is spherically symmetric and independent of $M$. 
When $J=0$, we omit $K$ and write $| v_1v_2v_3\rangle\equiv| v_1v_2v_30\rangle$. The basis states $|v_1v_2v_3K\rangle$ are eigenstates of $H_{\rm HO}$ but not of the full Hamiltonian $H$ because of the anharmonic and rovibrational coupling terms. We note that the eigenstates of $H_{\rm RR}$, the asymmetric top rotational wave functions, are superpositions of the  rotational basis states $|J,K\rangle$ \cite{Bauder2011}. The total angular momentum $J$ is a constant of motion, so  $J$ is not summed over in Eq.~\eqref{Eq:Psiexpansion}. The rovibrational wave functions for different values of $J$ are calculated independently. 

By selecting the $C_{2v}$ symmetry axis of H$_2$O to be the $c$ axis, the parity under proton exchange is given by 
\begin{equation}\label{Eq:ProtonExParityDef}
P_{\rm ex}=(-1)^{v_3+K}. 
\end{equation}
This means that states with even $v_3+K$ have proton spin $s=0$ (para-water), and states with odd $v_3+K$ have $s=1$ (ortho-water). The Hamiltonian $H$ does not couple states with different parity  $(-1)^{v_3+K}$.

We aim to obtain approximate eigenfunctions of $H$ by solving the eigenvalue problem
\begin{equation}
\bm{H}\bm{c}=E\bm{c},
\end{equation}
where $\bm{H}$ is the matrix representation of $H$ with matrix elements
\begin{equation}\label{Eq:HmatrixDef}
H_{v_1v_2v_3K,v_1'v_2'v_3'K'}=\langle v_1v_2v_3K|H|v_1'v_2'v_3'K'\rangle.
\end{equation}
For H$_2$O, the matrix $\bm{H}$ is of size $\Hsize\times\Hsize$, where $\Hsize=(\vmax+1)^3(2J+1)$, and is small enough to diagonalize on a classical computer using standard methods. 
The motivation to use a quantum computer for the eigenvalue evaluation is that for a molecule having  $\Nmodes$ vibrational modes, $\Hsize$ scales exponentially with $\Nmodes$ as  
$\Hsize=(\vmax+1)^{\Nmodes}(2J+1)$, which makes it difficult to calculate eigenvalues when $\Nmodes$ is large. In the present investigation, H$_2$O is selected as a test case where we can obtain the eigenvalues both on classical and quantum computers, so that the accuracy of the quantum algorithm can be assessed in a straightforward way.

The matrix representation $\bm{H}$ of the rovibrational Hamiltonian \eqref{Eq:HmatrixDef}  is a sparse, band-diagonal matrix. The matrix elements $H_{v_1v_2v_3K,v_1'v_2'v_3'K'}$ are non-zero only when 
\begin{equation}\label{Eq:Kselectionrule}
|K -K'|\le2
\end{equation}
and 
\begin{equation}\label{Eq:vselectionrule}
|v_k-v'_k|\le 4. %\text{ (for all $k$)}.
\end{equation} 
The reason for the inequality \eqref{Eq:Kselectionrule} is that the angular momentum operators $J_\alpha$ occur at most twice in $H$, and (for example) $\langle J,K|J_aJ_b|J,K'\rangle=0$ when $|K -K'|>2$.
%We have $\langle J,K|J_aJ_b|J,K+2\rangle=(\hbar^2/4i)\sqrt{J(J+1)-(K+2)(K+1)}\sqrt{J(J+1)-KK(+1)}$.
The vibrational selection rule \eqref{Eq:vselectionrule} derives from the fact that the mode coordinate $Q_k$ appears at most to fourth order in $H$, and $\langle v|Q^4|v'\rangle=0$ for 
$|v-v'|> 4$. The matrix elements of the operator $\bm{\pi}^T \bm{\mu}_\ell \bm{\pi}$ in Eq.~\eqref{Eq:HvibCorDef} also satisfy 
$\langle  v| \bm{\pi}^T \bm{\mu}_\ell \bm{\pi}|v'\rangle=0$ when $|v-v'|>4$ because $\ell\le 2$ and $\bm{\pi}^T \bm{\mu}_\ell \bm{\pi}$ therefore contains terms with $Q_k$ to the power of at most four. We also note that 
$\bm{\pi}$ does not contain terms $Q_kP_l$ with $k=l$ [see the definition in Eq.~\eqref{Eq:CorioloiscouplDef}].

%---------------------------------------------------------------------------------------------------------------------------------------------------------------------------------------------------------------
\subsection{Qubit form of the Hamiltonian}
%---------------------------------------------------------------------------------------------------------------------------------------------------------------------------------------------------------------
Next, we consider how the rovibrational structure problem can be implemented on a quantum computer. A standard introduction to many aspects of quantum computing can be found in Ref.~\onlinecite{NielsenChuang}. 
The first step is to derive  the qubit form of the Hamiltonian. There are several ways of encoding bosonic degrees of freedom as qubits \cite{Sawayaetal2020,Hanada2025,Mikkelsen2025}. Here, we use the binary mapping, in which a basis state is mapped to a qubit state according to
\begin{equation}\label{Eq:binarymapping}
|v_1v_2v_3K\rangle = |\bin(v_1)\bin(v_2)\bin(v_3)\bin(K)\rangle,
\end{equation} 
where $\bin(n)$ is the binary representation of the integer $n$. This representation requires $\nqubit=\nqubitvib+\nqubitrot=3\lceil\log_2(\vmax+1)\rceil+\lceil\log_2(2J+1)\rceil$ qubits, where $\lceil\cdot\rceil$ denotes the ceiling function. The binary encoding has the advantage of being compact in the sense of requiring a small number of qubits,  and can therefore be applied to very large molecules in principle. 

The qubit form of the Hamiltonian is given by an expansion
\begin{equation}\label{Eq:qubitHamiltonian}
H_{\text{q}} = \sum_{\bm{k}}h_{\bm{k}} P_{\bm{k}},
\end{equation}
where $h_{\bm{k}}$ is a real-valued coefficient, $\bm{k}=(k_0,k_1,\ldots,k_{\nqubit-1})$ is a composite index ($k_j\in\{0,1,2,3\}$), and
$P_{\bm{k}}$ is a direct product of Pauli matrices,
\begin{equation}\label{Eq:PaulistringDef}
P_{\bm{k}}=\sigma_{k_{\nqubit-1}}\otimes \cdots \otimes \sigma_{k_1} \otimes \sigma_{k_0},
\end{equation}
where $\sigma_0=I$, $\sigma_1=X$, $\sigma_2=Y$, and $\sigma_3=Z$.
We use $\NHq$ to denote the number of terms having non-zero $h_{\bm{k}}$ in \eqref{Eq:qubitHamiltonian}.

The numerical values of the $h_{\bm{k}}$ coefficients can be derived by expanding each term in Eq.~\eqref{Eq:Hdef} in terms of $Q_k$, $P_k$ and $J_\alpha$, converting each term to  an expression like \eqref{Eq:qubitHamiltonian}, and combining the expansions. An alternative way, employed in the present investigation,  is to evaluate $h_{\bm{k}}$ according to 
\begin{equation}\label{Eq:tracePH}
h_{\bm{k}}=\frac{1}{2^{\nqubit}}\trace (P_{\bm{k}}\bm{H}_{\rm q}),
\end{equation}
where the matrix elements of the $2^{\nqubit}\times 2^{\nqubit}$ matrix  $\bm{H}_{\rm q}$ equal those of $\bm{H}$ for all qubit states which represent rovibrational basis states, and equal zero otherwise.
The size of $\bm{H}_{\rm q}$ is larger than that of $\bm{H}$ if $\nqubit>3\log_2(\vmax+1)+\log_2(2J+1)$. Although \eqref{Eq:tracePH} is easy to use, a naive application requires the evaluation of $4^{\nqubit}$ expansion coefficients, which limits the number of qubits to $\nqubit<10$ or so \cite{Hantzko2024}.

%@@@@@@@@@@@@@@@@@@@@@@@@@@@@@@@@@@@@@@@@@@@@@@@@@@@@@@@@@@@@@@@@@@@@@@@@@@@@@@@@@@@@@@@
\begin{figure}
\includegraphics[width=\figwidth]{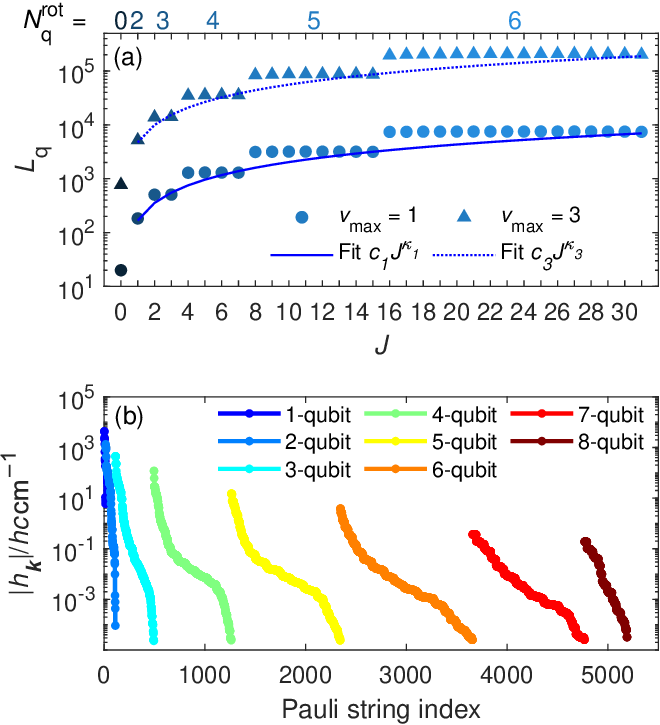}
\caption{\label{Fig1}(a) Number of terms $\NHq$ in the qubit Hamiltonian for $\vmax =1$ ({\large $\bullet$}) and $\vmax=3$ ($\blacktriangle$). The number of qubits $\nqubitrot$ used to represent the rotational part of the wave function is indicated above  the plot. The total number of qubits is $\nqubit=\nqubitrot+\nqubitvib=\nqubitrot+3$ for $\vmax=1$ and 
$\nqubit = \nqubitrot+6$ for $\vmax=3$.
The solid and dotted lines represent power-law  fits of the form $\NHq=c_{\vmax}J^{\kappa_{\vmax}}$, with $(c_1,\kappa_1)=(167.7,1.08)$ and $(c_3,\kappa_3)=(474,1.07)$. (b) Absolute values $|h_{\bm{k}}|$ of the expansion coefficients in the qubit Hamiltonian [see Eq.~\eqref{Eq:qubitHamiltonian}], for $J=1$. The coefficients $h_{\bm{k}}$ are sorted according to size and the number of qubits the Pauli operator $P_{\bm{k}}$ operates on.
}
\end{figure}
%@@@@@@@@@@@@@@@@@@@@@@@@@@@@@@@@@@@@@@@@@@@@@@@@@@@@@@@@@@@@@@@@@@@@@@@@@@@@@@@@@@@@@@@

We have derived the qubit form of the Watson Hamiltonian for H$_2$O for $\vmax =1$ and 3, and $J<32$. The number $\NHq$ of terms in the expansion \eqref{Eq:qubitHamiltonian} is displayed in Fig.~\ref{Fig1}(a). We can see that there are many terms in the qubit Hamiltonian. For $J=31$ and $\vmax =3$ ($\nqubit=12$), we have $\NHq = 201157$. Although large, $\NHq$ is still much smaller than the total number of possible $12$-qubit Pauli operators, $4^{12}\approx 1.7\times 10^7$.  We also see in Fig.~\ref{Fig1}(a) that $\NHq$ is approximately constant for values of $J$ resulting in the same number of $\nqubitrot$. This is a known property of the binary encoding \cite{Sawayaetal2020,Loetstedt2025}. The variation of 
$\NHq$ within the range of $J$ having the same $\nqubitrot$ is too small to be discernible in Fig.~\ref{Fig1}(a), but is at most a few percent of the average value. For example, for $\vmax=3$ and $16\le J\le 31$, corresponding to $\nqubitrot=6$, we have a minimum value of $\NHq^{\rm min}=195981$, a maximum value of $\NHq^{\rm max}=208184$, and an average value of $\overline{\NHq} \approx 204993$.

In order to get an idea of the scaling of $\NHq$ as a function of $J$, we make  a least-squares fit of $\NHq$ evaluated at $J=2^{\nqubitrot -1}-1$ (that is, the largest $J$ such that $2J+1<2^{\nqubitrot}$) to a power law of the form $\NHq = cJ^\kappa$. The fits are shown by solid and dotted curves in Fig.~\ref{Fig1}(a). For both $\vmax=1$ and $\vmax=3$, we find  $\kappa\approx 1$, showing that $\NHq$ is approximately proportional to $J$.

In Fig.~\ref{Fig1}(b), we show the absolute values $|h_{\bm{k}}|$ of the expansion coefficients for the example of $J=1$ for which $\nqubit =8$. We sort $|h_{\bm{k}}|$ according to size and the number of qubits the corresponding Pauli operator $P_{\bm{k}}$ operates non-trivially on. For example, if $\nqubit=4$, $P=IIXI$ is  a one-qubit operator, $P=IXYZ$ a three-qubit operator, and so on.  We can see in Fig.~\ref{Fig1}(b) that there are many terms in $H_{\rm q}$ operating on all eight qubits, but that these terms have small absolute values $|h_{\bm{k}}|<1$ \hcmminusone{}.

We also comment on the $\vmax$-dependence of $\NHq$. For $J=0$, we obtain $\NHq(\vmax=1)=20$, $\NHq(\vmax=3)=773$ [shown in Fig.~\ref{Fig1}(a)] and $\NHq(\vmax=7)=19491$ [not shown in Fig.~\ref{Fig1}(a)]. As we show in Appendix \ref{Appendix:vmaxscaling}, operators of the form $Q_1^{\ell_1}Q_2^{\ell_2}Q_3^{\ell_3}$ with $1\le \ell_j \le 4$ have qubit representations 
where the number of terms scale as 
$\bigO[\vmax^{3}\log_2^{3}(\vmax)]$. Therefore, we expect the   number of terms in $H_{\text{q}}$ to scale approximately as $\NHq=\bigO[J\vmax^{3}\log_2^{3}(\vmax)]$ for large $\vmax$. 
This scaling holds also for larger molecules with many vibrational modes, assuming that at most three-mode interactions are present in the Hamiltonian.

%---------------------------------------------------------------------------------------------------------------------------------------------------------------------------------------------------------------
\subsection{Quantum-selected configuration interaction}
%---------------------------------------------------------------------------------------------------------------------------------------------------------------------------------------------------------------
As can be seen in Fig.~\ref{Fig1}, the number of terms $\NHq$ in the qubit Hamiltonian is very large. The large value of $\NHq$ means that it is difficult to compute eigenstates by standard methods like the variation quantum eigensolver (VQE) \cite{Peruzzoetal2014,Yuan2019,Lietal2019,Parrishetal2019,Cerezo2021}. In the VQE method, we define an ansatz wave function $|\psi(\bm{\theta})\rangle$ depending on a number of variational parameters $\bm{\theta}=(\theta_1,\theta_2,\ldots)$, and minimize the expectation value of the Hamiltonian 
$\langle\psi(\bm{\theta})|H_{\rm q}|\psi(\bm{\theta})\rangle$ with respect to $\bm{\theta}$ to find the ground state wave function.  However, because each term in the qubit Hamiltonian has to be measured separately   in order to evaluate the expectation value of the Hamiltonian, a large number (of the order of $\NHq>10^3$ for H$_2$O) of circuits have to be executed and measured. Several methods have been proposed to reduce the number of  measurements \cite{Zhao2020, Verteletskyi2020, Majland2023, Yen2023,Reggio2024}. For example, in Ref.~\onlinecite{Majland2023},  a seven-fold reduction of the number of measurements  by optimizing the measurement scheme and coordinate transformations  was reported in vibrational VQE calculations of three-mode molecules. However, the number of required measurements remains large even after applying the reduction methods proposed in 
Refs.~\onlinecite{Zhao2020, Verteletskyi2020, Majland2023, Yen2023,Reggio2024}.
Taking into account that $\langle\psi(\bm{\theta})|H_{\rm q}|\psi(\bm{\theta})\rangle$ has to be evaluated many times in order to find the optimal value of 
$\bm{\theta}$, we conclude that it is not practically feasible to carry out  a VQE calculation on current quantum hardware. We comment that the problem of an excessive number of terms in the  qubit  Hamiltonian is not specific to rovibrational problems. Also  electronic structure qubit Hamiltonians have millions of terms in the expansion even for few-atomic molecules  (see for example Table II in Ref.~\onlinecite{Li2022}).

An alternative method for eigenvalue evaluation on quantum computers is the quantum-selected configuration interaction (QSCI) method \cite{Kanno2026,Nakagawa2024}, also referred to as the sample-based quantum diagonalization (SQD) method \cite{RobledoMoreno2025,Shirakawa2025,Shajan2025}. In the QSCI method,  a trial wave function  $|\psi(\bm{\theta})\rangle$ is prepared using a quantum computer, but unlike the VQE method, the energy is not directly evaluated from $|\psi(\bm{\theta})\rangle$. Instead, we measure $|\psi(\bm{\theta})\rangle$ in the computational basis $|x\rangle$ ($x= q_{\nqubit}q_{\nqubit-1}\cdots q_0$ with $q_j\in\{0,1\}$) to estimate the probability distribution
\begin{equation}
p(x)=|\langle x|\psi(\bm{\theta}) \rangle |^2.
\end{equation}
In contrast to the VQE method, only one circuit needs to be measured in the QSCI method.
Next, we select a subset $\Omega$ of basis states 
\begin{equation}
|x\rangle \in \{|x_1\rangle, |x_2\rangle,\ldots \}=\Omega
\end{equation}
based on the distribution $p(x)$. For example, we may select all $|x\rangle$ such that $p(x)>\epsilon$ for some  threshold value $\epsilon$. Approximate eigenstates of $H_{\rm q}$ are obtained by diagonalizing the Hamiltonian matrix $\bm{H}_\Omega$ expressed in the $\Omega$-basis, having matrix elements
\begin{equation}
H_{\Omega\,xx'}=\langle x|H_{\rm q}|x' \rangle,\qquad x,x'\in \Omega.
\end{equation}
The diagonalization is done using a classical computer. The role of the quantum computer is to provide an estimate of the basis functions contributing the most to an eigenstate.
The wave function is given in the $\Omega$-basis as 
\begin{equation}
|\phi\rangle =\sum_{x\in \Omega} c_x |x\rangle,
\end{equation}
where $\bm{c}=(c_{x_1}, c_{x_2},\ldots)^T$ is an eigenvector of $\bm{H}_\Omega$.
The number of basis states, $|\Omega|$, is assumed to be much smaller than the total number  $2^{\nqubit}$ of qubit states, and  to evaluate the  eigenvalues, the array of coefficients $\bm{c}$ must fit in the memory of the classical computer.
For this reason, the QSCI method is not expected to work for very large molecular systems.
If the basis set is large, complete diagonalization is not carried out; instead a few low-energy eigenstates are obtained by iterative methods such as the Davidson \cite{Davidson1975} and Lanczos \cite{50Lanczos,Golub2013}  methods. 
The QSCI method is expected to work well when an eigenstate of $H$ is well described using a small subset of states from the entire Hilbert state, as is commonly the case for electronic structure problems \cite{Xu2026}. As we shall see in Sec.~\ref{Subsec:QSCI}, the QSCI method works rather well also for rovibrational energy levels.

 We note that the QSCI method is not expected to be the best method for large, fault-tolerant quantum computers. In this case, 
we should use the quantum Krylov method 
\cite{StairHuangEvangelista2020,Cortes2022,Shen2023,Yoshioka2025} or the quantum phase estimation method \cite{NielsenChuang,Abrams1999,Aspuru-Guzik2005,Ollitrault2025,Yamamotoetal2026}. 

There are various proposals for how to construct the  trial wave function $|\psi(\bm{\theta}) \rangle$. 
Because the trial wave function does not need to accurately reproduce the target state in the QSCI method, we expect to be able to use a simpler quantum circuit for $|\psi(\bm{\theta}) \rangle$ than in the VQE method.
In Refs.~\onlinecite{RobledoMoreno2025,Shirakawa2025}, a unitary coupled-cluster-type circuit was used. In this paper, we employ a  trial wave function constructed by Hamiltonian time evolution \cite{Mikkelsen2025a,Sugisaki2025,Yu2025,Yamamoto2026},
\begin{equation}\label{Eq:TrotterexpitauH}
|\psi(\tau, \Ntrotter)\rangle = \UST^{\Ntrotter} |\psi_0\rangle,
\end{equation}
where $\Ntrotter$ is the number of Suzuki-Trotter steps, $|\psi_0\rangle$ is the initial wave function, and $\UST$ is a Suzuki-Trotter approximation \cite{Trotter1959,Suzuki1976,Ostmeyer2023} of the time-evolution operator, 
\begin{equation}\label{Eq:USTdef}
\UST=\prod_{|h_{\bm{k}}|>\lambda} e^{-i\tau h_{\bm{k}}P_{\bm{k}}/\hbar}.
\end{equation}
In Eq.~\eqref{Eq:USTdef},  we have introduced a cutoff parameter $\lambda$, meaning that in the time evolution operator $\UST$, we only include terms in the qubit Hamiltonian which have an expansion coefficient $h_{\bm{k}}$ with absolute value   larger than $\lambda$.
In the trial wave function $|\psi(\tau, \Ntrotter)\rangle$ defined in Eqs.~\eqref{Eq:TrotterexpitauH} and \eqref{Eq:USTdef}, the time step length $\tau$, the number of Suzuki-Trotter steps $\Ntrotter$ and the energy cutoff parameter $\lambda$ are  considered as free (variational) parameters which can be varied to lower the energy. The goal is to prepare a wave function 
$|\psi(\tau, \Ntrotter)\rangle$ which approximately is composed of the same set of  basis functions as the state we are trying to calculate (for example the ground state). The initial state  $|\psi_0\rangle$ in Eq.~\eqref{Eq:TrotterexpitauH} should be  easy to prepare on the quantum computer, and should have  a large overlap with the target state.

The time-evolution operator $\UST$ acting on $|\psi_0\rangle$ approximately generates a superposition of states of the form $H_{\rm q}^n|\psi_0\rangle$, meaning that  $|\psi(\tau, \Ntrotter)\rangle$ is composed of states which are coupled to $|\psi_0\rangle$ by powers of the Hamiltonian $H_{\rm q}$. Therefore, the basis states contained in $|\psi(\tau, \Ntrotter)\rangle$ should be good candidates for an accurate basis set for a target state $|\phi\rangle$ which has a large overlap with $|\psi_0\rangle$.

In the case of the evaluation of rovibrational eigenstates, we are almost always interested in obtaining many eigenstates, not just the ground state. We adopt the approach of selecting a different initial state $|\psi_0\rangle$ and sampling a different basis set for each excited state. The detailed procedure is given in Sec.~\ref{Subsec:QSCI}.

%=========================================================================================================================
\section{Results}\label{Sec:Results}
%=========================================================================================================================
In this section, we present numerical results of  the theory described in Sec.~\ref{Sec:Theory} applied  to  H$_2$O. The  details of our model of H$_2$O can be found in Appendix \ref{Appendix:H2Oparameters}. 

%---------------------------------------------------------------------------------------------------------------------------------------------------------------------------------------------------------------
\subsection{Energy spectrum and the importance of rovibrational coupling}\label{Subsec:Spectrum}
%---------------------------------------------------------------------------------------------------------------------------------------------------------------------------------------------------------------
Before the discussion on quantum computing, we start by investigating the spectrum of the rovibrational Hamiltonian. All results in this section \ref{Subsec:Spectrum} are obtained using standard, classical methods. In Table~\ref{Table1}, we show the  energy of the rovibrational ground state $E_{000}$ ($J=0$) and the vibrational band origins of the four lowest excited states, that is, the energy difference  between the state $v_1v_2v_3$ and the ground state defined as 
\begin{equation}\label{Eq:DefDeltaE}
\Delta E_{v_1v_2v_3}=E_{v_1v_2v_3}-E_{000}.
\end{equation}
 The label $v_1v_2v_3$ of an eigenstate $|\phi\rangle$ is decided by the basis state $|v_1v_2v_3\rangle$ having the largest overlap $|\langle v_1v_2v_3|\phi\rangle|$ with $|\phi\rangle$.  This is an unambiguous way of labeling eigenstates for small excitation energies. For large excitation energies above about 12000 \hcmminusone{}, straightforward labeling of the eigenstates becomes difficult because of the pronounced mixing of different basis functions \cite{Matyus2010}. 

In Table~\ref{Table1}, we show the vibrational energies obtained in the harmonic oscillator approximation,
\begin{equation}
E^{\text{HO}}_{v_1v_2v_3}=\hbar\sum_{k=1}^3 \left(\frac{1}{2}+v_k\right)\omega_k,
\end{equation}
by diagonalization of the matrix representation of the full rovibrational Hamiltonian $H$ using different values of the maximum vibrational quantum number $\vmax$, and by second-order perturbation theory. In second-order perturbation theory (PT2), we consider the harmonic-oscillator Hamiltonian $H_{\rm HO}$ to be the unperturbed Hamiltonian, and the rest of $H$ to be the perturbation $H'$, that is, $H=H_{\rm HO} + H'$. The PT2 vibrational energy is obtained as 
\begin{align}\label{Eq:PT2def}
E^{\text{PT2}}_{v_1v_2v_3}={}&E^{\text{HO}}_{v_1v_2v_3} + \langle v_1v_2v_3| H' |v_1v_2v_3\rangle 
\nonumber\\
&+ \sum_{\substack{v'_1,v'_2,v'_3=0\\v'_1v'_2v'_3\neq v_1v_2v_3}}^{\vmax} \frac{|\langle v_1v_2v_3 |H'|v'_1v'_2v'_3\rangle|^2}{E^{\text{HO}}_{v_1v_2v_3}-E^{\text{HO}}_{v'_1v'_2v'_3}}.
\end{align}
We mention that in the presence of anharmonic resonances [levels for which  the denominator $E^{\text{HO}}_{v_1v_2v_3}-E^{\text{HO}}_{v'_1v'_2v'_3}$ in \eqref{Eq:PT2def} is small for some $v'_1v'_2v'_3$], a special treatment is necessary in vibrational perturbation theory \cite{Amos1991, East1995, Csaszar1997, Ramakrishnan2015}. The five lowest states shown in Table~\ref{Table1} are well separated from other states, and Eq.~\eqref{Eq:PT2def} can therefore be used without modification.
For reference, we also show the experimentally measured vibrational energies from Ref.~\onlinecite{Furtenbacher2020a} in Table~\ref{Table1}.

%@@@@@@@@@@@@@@@@@@@@@@@@@@@@@@@@@@@@@@@@@@@@@@@@@@@@@@@@@@@@@@@@@@@@@@@@@@@@@@@@@@@@@@@
\begin{table}
\caption{\label{Table1}Energies in \hcmminusone{} of the vibrational ground state [column labeled by $000$], and the vibrational band origins 
$\Delta E_{v_1v_2v_3}=E_{v_1v_2v_3}-E_{000}$ [columns with $v_1v_2v_3\neq000$] for $J=0$, obtained using the harmonic oscillator approximation, the full rovibrational Hamiltonian $H$ with different values of $\vmax$, and second-order perturbation theory [PT2; see the definition in Eq.~\eqref{Eq:PT2def}]. In PT2, $\vmax=7$ is employed. We also show the  experimentally measured energies.}
\begin{ruledtabular}
\begin{tabular}{l|ddddd}
                                   &     0  0  0    &  0  1  0  &     0  2  0   &   1  0  0   &    0  0  1            \\
\hline
HO                                 &    4710.5     &   1649.7   &     3299.5    &    3830.9    &    3940.5            \\
$\vmax=3$                          &    4641.5     &   1590.2    &     3162.0    &    3715.8    &   3797.0             \\
$\vmax=5$                          &     4640.8    &   1588.4    &     3141.6    &    3688.3   &     3795.4            \\
$\vmax=7$                          &     4640.8    &    1588.4    &    3139.9    &    3687.8    &    3795.4             \\
PT2                                &     4628.7    &    1596.6     &    3153.4    &    3630.2    &    3732.3              \\
Exp.\footnote{W2020  database (Ref.~\onlinecite{Furtenbacher2020a}).}   
                                   &               &     1594.7     &   3151.6     &   3657.1    &    3755.9             \\
\end{tabular}
\end{ruledtabular}
\end{table}
%Data from Matlab output by Energylevels_H2O_nmaxconvergence.m
%(note that below the order is v2v1v3 ! (bending -- symstretch -- antisymstretch)
%Energies in cm-1
%                    0  0  0       1  0  0       2  0  0       0  1  0       0  0  1       
%HO                   4710.5        1649.7        3299.5        3830.9        3940.5
%Rovib(vmax=3)        4641.5        1590.2        3162.0        3715.8        3797.0
%Rovib(vmax=5)        4640.8        1588.4        3141.6        3688.3        3795.4
%Rovib(vmax=7)        4640.8        1588.4        3139.9        3687.8        3795.4
%Pert2(vmax=3)        4628.9        1597.9        3171.4        3655.2        3732.7
%Pert2(vmax=5)        4628.7        1596.6        3153.4        3630.2        3732.3
%Pert2(vmax=7)        4628.7        1596.6        3153.4        3630.2        3732.3
%Exp                                1594.7        3151.6        3657.1        3755.9
%@@@@@@@@@@@@@@@@@@@@@@@@@@@@@@@@@@@@@@@@@@@@@@@@@@@@@@@@@@@@@@@@@@@@@@@@@@@@@@@@@@@@@@@

We make several observations about the data in Table~\ref{Table1}. By comparing the harmonic oscillator energies and the energies obtained using $H$, we see that the anharmonic couplings can shift the vibrational energies with more than 100 \hcmminusone{}. Second, the vibrational energies can be considered to be converged at $\vmax =7$, although vibrational energies accurate to within about 30 \hcmminusone{} can be obtained already at $\vmax =3$. Third, PT2 is not good enough for obtaining accurate eigenenergies. The difference between the $\Delta E_{v_1v_2v_3}$ obtained using PT2 ($\vmax=7$) and diagonalization of $\bm{H}$ ($\vmax=7$) is larger than 50 \hcmminusone{} for $\Delta E_{100}$ and $\Delta E_{001}$.  

Finally, we note that the difference between the 
excitation energies obtained in our model  of H$_2$O ($\vmax=7$) and the experimentally measured excitation energies is as large as about  40 \hcmminusone{} for $\Delta E_{001}$.
 In order to  reproduce the experimental vibrational energies to within 1 \hcmminusone{}, improved potential energy surfaces combined with more sophisticated basis sets (such as those based on a discrete-variable representation) are required \cite{Czako2004,Matyus2007}. Curiously, the PT2  energies actually agree better with  the experimental values  than the energies obtained by diagonalization of $\bm{H}$ \cite{Csaszar1997}, even though there is still a discrepancy between  PT2  and experiment  of  more than 20 \hcmminusone{} for $\Delta E_{100}$ and $\Delta E_{001}$.

%@@@@@@@@@@@@@@@@@@@@@@@@@@@@@@@@@@@@@@@@@@@@@@@@@@@@@@@@@@@@@@@@@@@@@@@@@@@@@@@@@@@@@@@
\begin{figure}
\includegraphics[width=\figwidth]{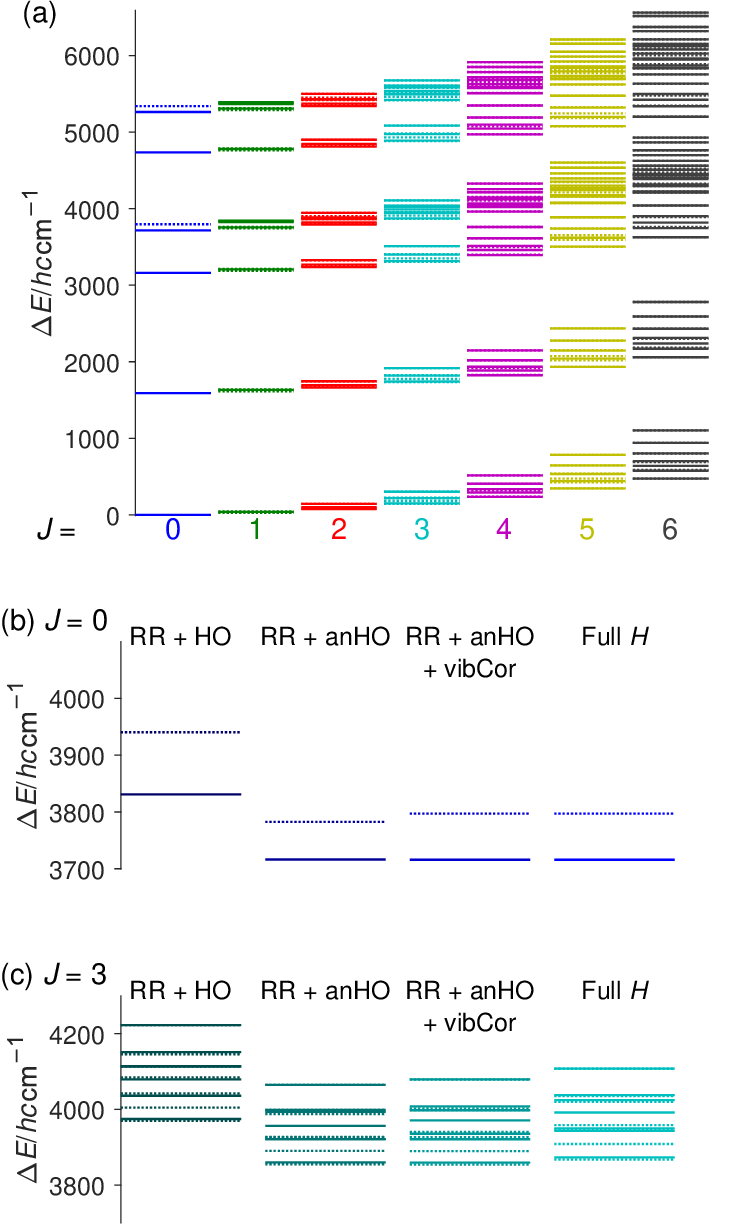}
\caption{\label{Fig2}(a) Rovibrational energy levels $\Delta E$ relative to the rovibrational ground state for $0\le J \le 6$ obtained using the full rovibrational Hamiltonian defined in Eq.~\eqref{Eq:Hdef}. The maximum vibrational quantum number is $\vmax=3$. Energy levels of proton singlet states are drawn by solid lines and triplet states are drawn by dotted lines.
(b) Energy spectrum in the range 
$3700\text{ \hcmminusone{}}\le\Delta E\le 4000\text{ \hcmminusone{}}$ for $J=0$. We compare the energy levels obtained using the Hamiltonians $H_{\rm RR} + H_{\rm HO}$ (first  column), 
$H_{\rm RR} + H_{\rm HO} + V_{\rm anharm}$ (second column), $H_{\rm RR} + H_{\rm HO} + V_{\rm anharm} + H_{\rm vibCor}$ (third column), and $H$ (fourth column). See Eqs.~\eqref{Eq:Hdef}--\eqref{Eq:Hrovibdef} for definitions of the different Hamiltonians.
(c) Energy spectrum in the range 
$3800\text{ \hcmminusone{}}\le\Delta E\le 4300\text{ \hcmminusone{}}$ for $J=3$. 
}
\end{figure}
%@@@@@@@@@@@@@@@@@@@@@@@@@@@@@@@@@@@@@@@@@@@@@@@@@@@@@@@@@@@@@@@@@@@@@@@@@@@@@@@@@@@@@@@

In Fig.~\ref{Fig2}(a), we show the rovibrational energy spectrum $\Delta E$ for $\Delta E<6600$ \hcmminusone{} and  $J$ up to 6, obtained by diagonalization of $\bm{H}$. The energy $\Delta E$ is defined as the excitation energy relative to the rovibrational ground state (the lowest energy obtained at $J=0$). For large values of $J$, the density of states becomes large with many close-lying states. For example, for $J=6$ and $\Delta E$ around 5000 \hcmminusone{}, the energy gap between neighboring levels  is typically around 10 \hcmminusone{}.

In order to illustrate the importance of the different terms in the Hamiltonian \eqref{Eq:Hdef}, we show in  Fig.~\ref{Fig2}(b) and (c) the energy levels around 4000 \hcmminusone{} for $J=0$ and $3$ obtained at different levels of approximation. We compare the energy levels obtained using $H_{\rm RR} + H_{\rm HO}$ (harmonic oscillator approximation), $H_{\rm RR} + H_{\rm HO} + V_{\rm anharm}$ (anharmonic potential energy surface, but no rovibrational coupling), $H_{\rm RR} + H_{\rm HO} + V_{\rm anharm} + H_{\rm vibCor}$ (including also vibrational Coriolis coupling), and the full $H$ (including rovibrational coupling). We can see in Fig.~\ref{Fig2}(b) and (c) that even if the largest energy change ($\sim 100$ \hcmminusone{}) is due to the inclusion of the anharmonic terms $V_{\rm anharm}$ in the Hamiltonian, the vibrational Coriolis coupling $H_{\rm vibCor}$ and the rovibrational coupling $H_{\rm  rovib}$ are crucial for obtaining accurate results. For example, as can be seen in Fig.~\ref{Fig2}(b) by comparing the ``RR + anHO'' and ``RR + anHO + vibCor'' columns, even in the rotational ground state ($J=0$), the $001$ level is shifted from $\Delta E_{001}=3782.6$ \hcmminusone{} by about 
14 \hcmminusone{} to $3797.0$ \hcmminusone{} by the inclusion of the vibrational Coriolis coupling term. Note that for $J=0$, $H_{\rm  rovib}=0$, which means that the ``Full $H$'' energies are the same as the ``RR + anHO + vibCor'' energies.   By comparing the ``RR + anHO + \mbox{vibCor}'' and the ```Full $H$'' columns in Fig.~\ref{Fig2}(c),
we can see that for rotating H$_2$O ($J=3$), the energy levels around 4000 \hcmminusone{} are shifted by up to 30 \hcmminusone{} by the addition of the rovibrational coupling term  $H_{\rm  rovib}$. Moreover, the order of the singlet and triplet states are changed for some states.

%---------------------------------------------------------------------------------------------------------------------------------------------------------------------------------------------------------------
\subsection{Distributions by quantum computers}\label{Subsec:Distributions}
%---------------------------------------------------------------------------------------------------------------------------------------------------------------------------------------------------------------

In this section and the next, we describe the results of QSCI calculations. We execute the quantum circuits defined in Eqs.~\eqref{Eq:TrotterexpitauH} and \eqref{Eq:USTdef} on Quantinuum's \Reimei{} quantum computer \cite{QuantinuumReimei2025}. \Reimei{} is a 20-qubit trapped-ion quantum  computer \cite{Pino2021} physically located on the Wako campus of RIKEN, Japan, and  features an all-to-all qubit connectivity and a two-qubit gate error of  around $\varepsilon_{\rm 2q}=1\times 10^{-3}$. Detailed values of the error rates of \Reimei{} at the time of the simulations are given in Appendix \ref{Appendix:ReimeiParameters}. The quantum circuits corresponding to $|\psi(\tau,\Ntrotter)\rangle$ are constructed and transpiled using Quantinuum {\sc nexus} \cite{quantinuum_nexus}
 and {\sc tket} \cite{Sivarajah2020}.

%@@@@@@@@@@@@@@@@@@@@@@@@@@@@@@@@@@@@@@@@@@@@@@@@@@@@@@@@@@@@@@@@@@@@@@@@@@@@@@@@@@@@@@@
\begin{figure*}
\includegraphics[width=\figwidthbroad]{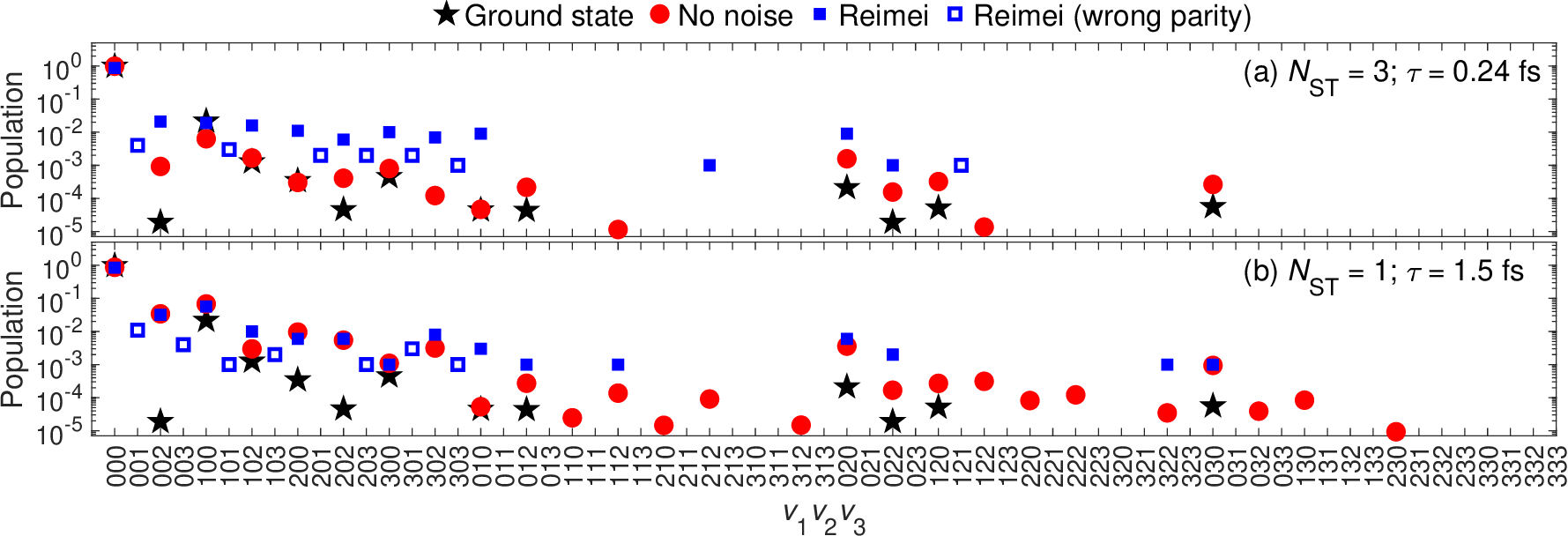}
\caption{\label{Fig3}
Distributions of basis states $|\langle v_1v_2v_3|\psi(\tau,\Ntrotter)\rangle|^2$ for (a) $\Ntrotter=3$, $\tau=0.24$ fs, and (b) $\Ntrotter=1$, $\tau=1.5$ fs. The exact, noise-free distribution is shown with red, filled circles, and the distribution estimated using \Reimei{} ($\Nshots=10^3$) is shown with blue squares. The states having an incorrect value of the parity $(-1)^{v_3}$ are displayed by open squares. The  populations $|c_{v_1v_2v_3}^0|^2$ in the  vibrational ground state are shown by black stars as a reference. 
}
\end{figure*}
%@@@@@@@@@@@@@@@@@@@@@@@@@@@@@@@@@@@@@@@@@@@@@@@@@@@@@@@@@@@@@@@@@@@@@@@@@@@@@@@@@@@@@@@

 In Fig.~\ref{Fig3}, we show two examples of the distribution  obtained by measuring the trial wave function $|\psi(\tau,\Ntrotter)\rangle$. The total angular momentum is $J=0$, 
 $\vmax =3$,  and the initial state is $|\psi_0\rangle = |v_1=0,v_2= 0,v_3= 0\rangle$ (corresponding to the all-zero qubit state). The total number of qubits employed is $\nqubit = 6$. The cutoff parameter $\lambda$ in Eq.~\eqref{Eq:USTdef} is taken to be $\lambda =110$ \hcmminusone{}. There are  a total of 40 terms (11 single-qubit, 17 two-qubit, 10 three-qubit, and two four-qubit)  in the qubit Hamiltonian \eqref{Eq:qubitHamiltonian} having $|h_{\bm{k}}|>110$ \hcmminusone{}. In Fig.~\ref{Fig3}(a), we show the result of a simulation 
 run on \Reimei{} on July 18, 2025 %jobname: QSCIrovib_J0_nmax3_000_TrotterCircuits_optlvl3_reimei
 employing $\Ntrotter =3$ and $\tau = 0.24$ fs ($= 10 \hbar/E_{\rm h}$, where $\hbar/E_{\rm h}$ is the atomic unit of time, and $\hbar = h/2\pi$) and in Fig.~\ref{Fig3}(b), we show the result of a simulation 
 run on \Reimei{} on November 18, 2025 %jobname: QSCIrovib_J0_nmax3_000_optlvl3_dtloopReimei_reimei
 with $\Ntrotter = 1$ and $\tau = 1.5$ fs ($=60\hbar/E_{\rm h}$).
 The transpiled circuit for $|\psi(\Ntrotter =3,\tau = 0.24\text{ fs})\rangle$ contains $85$ $R_{ZZ}$ two-qubit gates  and 134 single-qubit gates.
 The transpiled circuit for $|\psi(\Ntrotter =1,\tau = 1.5\text{ fs})\rangle$ contains $36$ $R_{ZZ}$ gates and 60 single-qubit gates. 
 Three distributions are displayed as a function of the basis function index $v_1v_2v_3$: (i) the probability distribution of the exact vibrational ground state 
 \begin{equation}
 p_{0}(v_1v_2v_3)=|c_{v_1v_2v_3}^0|^2,
 \end{equation}
 where $\bm{c}^0$ is the lowest-energy eigenvector of $\bm{H}$, (ii) the exact Suzuki-Trotter distribution 
 \begin{equation}\label{Eq:pSTDef}
 p_{\rm ST}(v_1v_2v_3)=|\langle v_1v_2v_3|\psi(\tau,\Ntrotter)\rangle|^2,
 \end{equation}
where $|\psi(\tau,\Ntrotter)\rangle$ is defined in Eqs.~\eqref{Eq:TrotterexpitauH} and \eqref{Eq:USTdef}, and (iii) the noisy distribution
\begin{equation}\label{Eq:pSTnoisyDef}
 p_{\rm noisy}(v_1v_2v_3)=|\langle v_1v_2v_3|\psi_{\rm noisy}(\tau,\Ntrotter)\rangle|^2,
\end{equation}
which is estimated by executing the circuit corresponding to $|\psi(\tau,\Ntrotter)\rangle$ using \Reimei{}. We measure the quantum state $\Nshots=10^3$ times in order to estimate 
\begin{equation}
p_{\rm noisy}(v_1v_2v_3)\approx\frac{N_{v_1v_2v_3}}{\Nshots},
\end{equation}
where $N_{v_1v_2v_3}$ is the number of measurements of $|v_1v_2v_3\rangle$. Error mitigation is not applied.

We can see in Figs.~\ref{Fig3}(a) and (b) that both the exact distribution $p_{\rm ST}$ and the noisy distribution  $p_{\rm noisy}$ are non-zero for basis states which have a non-zero population in the vibrational ground state, which corroborates the idea that the distributions can be used for generation of compact basis sets. The noisy distribution $p_{\rm noisy}$ agrees better with $p_{\rm ST}$ in Fig.~\ref{Fig3}(b) than in  Fig.~\ref{Fig3}(a) because of the shorter circuit and the smaller number of noisy two-qubit gates. Because of the finite number of shots ($\Nshots=10^3$) employed in the simulations run on \Reimei, some states having small populations in the exact Suzuki-Trotter distribution $p_{\rm ST}$ are not present in the \Reimei{} distribution $p_{\rm noisy}$.
The Suzuki-Trotter operator $\UST$  does not  couple  basis functions having different parity $(-1)^{v_3}$, and therefore basis states where $v_3$ is odd should not appear  in the distribution. However, on \Reimei,   we measure population in states with odd $v_3$ (shown with open squares in Fig.~\ref{Fig3}) due to the noise. When we perform the selection of the basis set by  sampling the distribution measured on \Reimei,  a simple error mitigation procedure consists of post-selecting only the basis functions having the correct parity, and renormalizing the distribution
\cite{Yoshida2022_copy,Cai2023,Ato2025},
\begin{equation}\label{Eq:pmitDef}
 p_{\rm mit}(v_1v_2v_3)= \frac{p_{\rm noisy}(v_1v_2v_3)}{\tilde{p}},
 \end{equation}
 where the normalization factor
 \begin{equation}
\tilde{p}= \sum_{(-1)^{v_3}=(-1)^{v_3^0}} p_{\rm noisy}(v_1v_2v_3),
\end{equation}
and $v_3^0$ is the initial-state vibrational quantum number of mode three.

%---------------------------------------------------------------------------------------------------------------------------------------------------------------------------------------------------------------
\subsection{Rovibrational energy levels by QSCI}\label{Subsec:QSCI}
%---------------------------------------------------------------------------------------------------------------------------------------------------------------------------------------------------------------

In order to evaluate the rovibrational energy levels, we sample from the distributions $p_{\rm ST}(v_1v_2v_3)$ and $p_{\rm mit}(v_1v_2v_3)$. We consider two cases: (i) a fixed value of the time step $\tau = 10 \hbar/E_{\rm h}$ and varying number of Suzuki-Trotter steps $0\le \Ntrotter\le 10$, and (ii) fixed $\Ntrotter=1$ and varying $\tau/(\hbar E_{\rm h}^{-1})=0, 20, 40, \ldots, 180$. 
 For each value of $\Ntrotter$ or $\tau$, we collect all basis functions  having a population $p_{\rm mit}(v_1v_2v_3)$ or $p_{\rm ST}(v_1v_2v_3)$ larger than a threshold value $\epsilon =10^{-4}$. We keep basis functions sampled at previous values of $\Ntrotter$ or $\tau$, and add newly sampled basis functions to the basis set $\Omega$. The size $|\Omega|$ of the basis set increases with $\Ntrotter$ in case (i), and with $\tau$ in case (ii). 
 For the simulations done using \Reimei{}, we employ $\Nshots=10^3$. This means that all  basis states $|v_1v_2v_3\rangle$ with non-zero $p_{\rm mit}(v_1v_2v_3)$ are added to the basis set,  because $\epsilon< 1/\Nshots$.
As we shall see in Secs.~\Ref{Subsubec:vibenerglvlsvmax3J0}--\Ref{Subsubec:vmax3J1}, both approaches give  vibrational energies converging to the exact energies. However, as approach (ii) results in shorter quantum circuits and therefore a smaller error due to the noise, approach (ii) may be said to be the best approach for near-term noisy quantum computers.

 With the idea of generating a separate,  optimal basis set for each excited state,  we generate five basis sets by changing the initial state $|\psi_0\rangle$, 
 \begin{equation}
 |\psi_0 (v_1v_2v_3)\rangle= |v_1v_2v_3\rangle
 \end{equation}
 for 
 \begin{equation}\label{Eq:initialstatespsi0}
 v_1v_2v_3\in \Gamma=\{000, \text{ } 010, \text{ } 020, \text{ }100, \text{ } 001\}.
 \end{equation}
 The set $\Gamma$ represents the five lowest vibrational states.
We label the basis set obtained (at certain values of $\Ntrotter$ and $\tau$) using an initial state $|\psi_0(v_1v_2v_3)\rangle$ by $\Omega_{v_1v_2v_3}$,
\begin{equation}
\Omega_{v_1v_2v_3}=\{|v_1v_2v_3\rangle,|v'_1v'_2v'_3\rangle, |v''_1v''_2v''_3\rangle, \ldots \}.
\end{equation}
Because the initial state for the generation of $\Omega_{v_1v_2v_3}$ is $|v_1v_2v_3\rangle$, 
$\Omega_{v_1v_2v_3}$ always contains $|v_1v_2v_3\rangle$, but the other basis functions can be different for each $\Omega_{v_1v_2v_3}$. 
 For each $\Omega_{v_1v_2v_3}$, eigenstates $|\phi_1\rangle,\ldots,|\phi_{|\Omega_{v_1v_2v_3}|}\rangle$ are obtained by diagonalization of the Hamiltonian matrix $\bm{H}_{\Omega_{v_1v_2v_3} }$.

The estimates of the energy levels are obtained by combining the five basis sets according to  the following procedure. 
From the set of eigenstates generated from the basis set $\Omega_{v_1v_2v_3}$, we select the eigenstate $|\phi_{v_1v_2v_3}\rangle$ having the largest overlap with the initial state 
$|v_1v_2v_3\rangle$. This results in a set of five states $\{|\phi_{v_1v_2v_3}\rangle\}$ with $v_1v_2v_3$ in the range defined in Eq.~\eqref{Eq:initialstatespsi0}. An orthonormal basis set 
$\{|\tilde{\phi}_{v_1v_2v_3}\rangle\}$ is produced from $\{|\phi_{v_1v_2v_3}\rangle\}$ by canonical orthogonalization \cite{Lowdin1970}, and the final estimates of the energy levels are obtained by diagonalization of  the $5\times 5$ Hamiltonian matrix with matrix elements $\langle \tilde{\phi}_{v_1v_2v_3}|H|\tilde{\phi}_{v'_1v'_2v'_3}\rangle$. In practice, we find that the absolute values of the off-diagonal elements of the overlap matrix $\langle\phi_{v_1v_2v_3}|\phi_{v'_1v'_2v'_3}\rangle$ are smaller than $0.05$ for both the basis sets sampled without noise and the basis sets sampled using \Reimei{}, meaning that the original states $\{|\phi_{v_1v_2v_3}\rangle\}$ are close to orthogonal.

We may also consider simply combining the five basis sets $\Omega_{v_1v_2v_3}$ into one large basis set 
\begin{equation}\label{Eq:Omegabig}
    \Omega_{\text{big}}=\bigcup_{v_1v_2v_3\in\Gamma}\Omega_{v_1v_2v_3}
\end{equation}
and computing the energy levels by diagonalization of the Hamiltonian matrix constructed from the basis functions in $\Omega_{\text{big}}$. However, because the basis functions in each $\Omega_{v_1v_2v_3}$ are different, the size of $\Omega_{\text{big}}$ is typically several times larger than the size of each $\Omega_{v_1v_2v_3}$. As exemplified  in 
Sec.~\ref{Subsubec:vibenerglvlsvmax3J0}, using $\Omega_{\text{big}}$ therefore results in less accurate energy estimates at the same basis set size $|\Omega_{\text{big}}|\approx |\Omega_{v_1v_2v_3}|$. 

We comment that the size of the set $\Gamma$ is limited only by practical considerations, such as the limited time available on the quantum computer. A larger number of excited states can be included in $\Gamma$ in a straightforward way by increasing the number of executed quantum circuits. One set of circuits is required for each state included in $\Gamma$.

We compare the energy levels obtained using quantum computers with the energy levels obtained using two classical methods. In the first method, we calculate the wave function in first-order perturbation theory, 
\begin{equation}
|\psi_{\rm PT}(v_1v_2v_3)\rangle= |v_1v_2v_3\rangle + |\delta \psi(v_1v_2v_3)\rangle,
\end{equation}
where
\begin{equation}\label{Eq:deltapsiPT}
|\delta \psi(v_1v_2v_3)\rangle=  \sum_{\substack{v'_1,v'_2,v'_3=0\\v'_1v'_2v'_3\neq v_1v_2v_3}}^{\vmax} \frac{\langle v'_1v'_2v'_3 |H'|v_1v_2v_3\rangle}{E^{\text{HO}}_{v_1v_2v_3}-E^{\text{HO}}_{v'_1v'_2v'_3}}|v'_1v'_2v'_3\rangle
\end{equation}
and the perturbation Hamiltonian $H'$ is defined as $H'=H-H_{\rm HO}$. The energy levels are obtained by sampling basis functions from the distribution given by the perturbative wave function \eqref{Eq:deltapsiPT},
\begin{equation}\label{Eq:pPTDef}
p_{\rm PT}(v_1v_2v_3) = |\langle v_1v_2v_3|\psi_{\rm PT}(v_1v_2v_3)\rangle|^2,
\end{equation}
using the same threshold value $\epsilon = 10^{-4}$ as for the distributions obtained using quantum computers, and diagonalizing the Hamiltonian matrix. Because of the final diagonalization step, this procedure results in more accurate energy estimates than the second-order perturbation theory \eqref{Eq:PT2def}. Moreover, this method is in principle scalable to large systems with many vibrational modes if the sum in \eqref{Eq:deltapsiPT} is truncated to include only terms with large values of  
$|\langle v'_1v'_2v'_3 |H'|v_1v_2v_3\rangle/(E^{\text{HO}}_{v_1v_2v_3}-E^{\text{HO}}_{v'_1v'_2v'_3})|$. We note that similar methods of selecting the dominant set of basis  functions have been proposed 
in electronic structure theory, two representative examples being the
configuration interaction by perturbation with multiconfigurational 
zeroth-order wave function selected by iterative process (CIPSI) method \cite{Huron1973} and the heat-bath CI method \cite{Holmes2016}.  For the simulation of vibrational structure, the vibrational heat-bath CI method \cite{Fetherolf2021, Bhatty2021, Tran2023}
has been developed.

In the second method, referred to in the following as the ``optimal`` method, we aim to select the best possible basis set for each state for a given size $|\Omega|$ of the basis set. We determine the basis sets by adding basis functions one by one  as follows. 
We start at $|\Omega|=1$, for which the optimal basis sets for all states in $\Gamma$ contain only one basis function,
$\Omega_{v_1v_2v_3}^{\text{opt}} =\{|v_1v_2v_3\rangle \}$. For $|\Omega|=2$, we first determine the optimal basis set for the vibrational ground state, $v_1v_2v_3=000$. 
We create a basis set with two basis functions by adding the basis function $|v'_1v'_2v'_3\rangle$ that minimizes the energy of the state $|\varphi_{000}\rangle$, where $|\varphi_{000}\rangle$ is defined as the state having the largest overlap with $|000\rangle$. In practice, this means that we test all basis sets $\{|000\rangle, |v'_1v'_2v'_3\rangle \}$ with $v'_1v'_2v'_3\neq 000$, diagonalize the $2\times 2$ Hamiltonian matrix and find the basis set that gives the minimum energy of $|\varphi_{000}\rangle$. 
Next, we generate an $|\Omega|=2$ basis set for the next state in $\Gamma$, $v_1v_2v_3=010$. We take a basis function $|v'_1v'_2v'_3\rangle$, diagonalize the Hamiltonian matrix in the $\{|010\rangle, |v'_1v'_2v'_3\rangle \}$  basis set, and obtain
$|\varphi_{010}\rangle$. We then calculate the energy of $|\varphi_{010}\rangle$ by diagonalizing the Hamiltonian matrix constructed from the (orthogonalized)
$\{|\varphi_{000}\rangle,|\varphi_{010}\rangle\}$ basis set. We repeat for all $|v'_1v'_2v'_3\rangle\ne |010\rangle$ 
and select the basis function $|v'_1v'_2v'_3\rangle$ which minimizes the energy. For the remaining states in $\Gamma$, the procedure is similar. For example, for 
 $v_1v_2v_3=100$, we select the second basis state of the $\Omega_{100}^{\text{opt}}$ basis set so that the $100$ energy of the Hamiltonian constructed from the $\{|\varphi_{000}\rangle,|\varphi_{010}\rangle, |\varphi_{020}\rangle,|\varphi_{100}\rangle\}$ basis set is minimized. The final  estimates for energy levels are obtained by the complete combined basis set 
 $\{|\varphi_{000}\rangle,\ldots,|\varphi_{001}\rangle\}$.
 To construct basis sets with three basis functions ($|\Omega|=3$) for a state $v_1v_2v_3$, we add one basis function $|v''_1v''_2v''_3\rangle$ to the basis set $\{|v_1v_2v_3\rangle, |v'_1v'_2v'_3\rangle \}$ obtained for $|\Omega|=2$. To select $|v''_1v''_2v''_3\rangle$, we calculate 
 $|\varphi_{v_1v_2v_3}\rangle$ by diagonalizing the $3\times3$ Hamiltonian in the $\{|v_1v_2v_3\rangle, |v'_1v'_2v'_3\rangle, |v''_1v''_2v''_3\rangle \}$ basis, calculate the energy from the combined basis set 
 $\{|\varphi_{000}\rangle,\ldots,|\varphi_{v_1v_2v_3}\rangle\}$ (each $|\varphi_{v_1v_2v_3}\rangle$ is a superposition of three basis functions), and select the basis function $|v''_1v''_2v''_3\rangle$ that minimizes the energy.
 
 The above procedure is repeated for all basis set sizes up to  $|\Omega|=(\vmax +1)^3/2$, which is the maximum number of basis functions of a given parity. At a certain value of $|\Omega|$ and state $v_1v_2v_3$, the basis function added to the basis set is determined by the minimization of the energy given by the combined basis set  
 $\{|\varphi_{000}\rangle,\ldots,|\varphi_{v_1v_2v_3}\rangle\}$. 
By minimizing the energy obtained by the combined basis set, we obtain energy levels which are guaranteed to converge to the exact energy levels at large $|\Omega|$.

%@@@@@@@@@@@@@@@@@@@@@@@@@@@@@@@@@@@@@@@@@@@@@@@@@@@@@@@@@@@@@@@@@@@@@@@@@@@@@@@@@@@@@@@
\begin{figure}
\includegraphics[width=\figwidth]{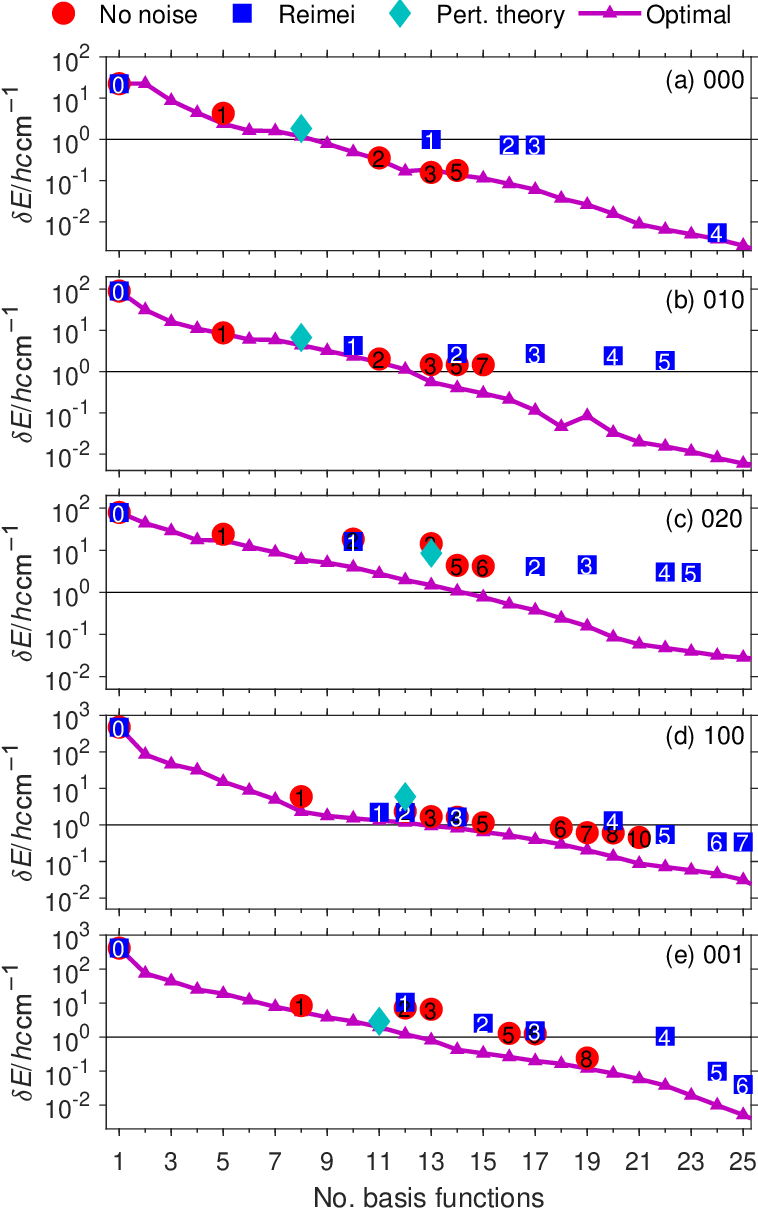}
\caption{\label{Fig4}
Energy differences $\delta E_{v_1v_2v_3}$ on a logarithmic scale for $v_1v_2v_3$ equal to (a) 000, (b) 010, (c) 020, (d) 100, and (e) 001, as a function of the number of basis functions in the basis set. $J=0$ and $\vmax=3$ (corresponding to a number of qubits $\nqubit =6$).  The   values of $\delta E_{v_1v_2v_3}$ obtained by sampling of the 
Suzuki-Trotter distribution \eqref{Eq:pmitDef} are shown with filled red circles, those obtained by sampling the mitigated distribution \eqref{Eq:pmitDef} obtained using \Reimei{} are shown with blue, filled squares. The time step size is fixed to $\tau=0.24$ fs [same as in Fig.~\ref{Fig3}(a)], and the number of Suzuki-Trotter steps varies from $\Ntrotter =0$ to $\Ntrotter =7$. 
At $\Ntrotter =0$, the basis set contains   only the initial state $|v_1v_2v_3\rangle$. 
The value of $\Ntrotter$ is indicated inside the curve symbols.
The diamond symbol shows $\delta E_{v_1v_2v_3}$ obtained by sampling from the perturbation-theory distribution \eqref{Eq:pPTDef}, and the $\delta E_{v_1v_2v_3}$'s obtained by the optimal method are shown with purple triangles.
The energy difference of 1 \hcmminusone{} is indicated by the black horizontal line.
}
\end{figure}
%Data taken on Reimei during July 18, 23, 24, 25, 26
%@@@@@@@@@@@@@@@@@@@@@@@@@@@@@@@@@@@@@@@@@@@@@@@@@@@@@@@@@@@@@@@@@@@@@@@@@@@@@@@@@@@@@@@

%...............................................................................................................................................................................................................................................................................................................................................................................................
\subsubsection{ 
Vibrational energy levels for $\vmax=3$ and $J=0$}\label{Subsubec:vibenerglvlsvmax3J0}
%...............................................................................................................................................................................................................................................................................................................................................................................................
In Fig.~\ref{Fig4}, we show the results of rovibrational QSCI simulations carried out on \Reimei{} during July 18--26, 2025. The model parameters employed are $J=0$ and $\vmax=3$ (implying a total number of qubits $\nqubit =6$), $\tau = 0.24$ fs ($=10\hbar/E_{\rm h}$), and a varying number of Suzuki-Trotter steps $\Ntrotter$. We consider the energy levels for the eigenstates labeled by $v_1v_2v_3$ in $\Gamma$ [see Eq.~\eqref{Eq:initialstatespsi0}], obtained by sampling the distributions \eqref{Eq:pSTDef} and  \eqref{Eq:pmitDef}. For comparison, we also show results obtained by sampling the perturbation-theory distribution \eqref{Eq:pPTDef}, and the energies obtained by the optimal method. 
In Fig.~\ref{Fig4}, we show the energy difference $\delta E_{v_1v_2v_3}$ defined relative to the exact energy,
\begin{equation}\label{Eq:smalldeltaEDef}
\delta E_{v_1v_2v_3} = E_{v_1v_2v_3}-E_{v_1v_2v_3}^{\rm exact},
\end{equation}
where $E_{v_1v_2v_3}^{\rm exact}$ is the exact eigenvalue of the full matrix $\bm{H}$ including all $(\vmax+1)^3=64$ basis functions (for numerical values, see Table~\ref{Table1}). Note that $\delta E_{v_1v_2v_3}$ defined in \eqref{Eq:smalldeltaEDef} is different from $\Delta E_{v_1v_2v_3}$ defined in \eqref{Eq:DefDeltaE}.

We can see in Fig.~\ref{Fig4} that in the noise-free case (filled red circles), $\delta E_{v_1v_2v_3}$ converges to a small value  in all cases shown, and becomes close to $\delta E_{v_1v_2v_3}$ obtained using the optimal method at about $\Ntrotter =5$ Suzuki-Trotter steps. After one Suzuki-Trotter step ($\Ntrotter =1$), typically five to ten basis functions are sampled. If we consider the largest basis set generated by the sampling (the rightmost red circle in each panel), $\delta E_{v_1v_2v_3}$ is smaller than 1 \hcmminusone{}  in three cases [(a), (d), and (e)], while we have 
$\delta E_{010}^{\text{no noise}}(\Ntrotter = 7)  \approx 1.5$ \hcmminusone{} in Fig.~\ref{Fig4}(b) and
$\delta E_{020}^{\text{no noise}}(\Ntrotter = 6) \approx 4.2$ \hcmminusone{} in Fig.~\ref{Fig4}(c).

Because of the noise, the distribution $p_{\rm mit}(v_1v_2v_3)$ measured on  \Reimei{} is different from the noise-free distribution, as can be seen in Fig.~\ref{Fig3}. The basis sets sampled on \Reimei{} are larger than in the noise-free case. Typically, 10 to 12 basis functions are sampled after one Suzuki-Trotter step on \Reimei. Nevertheless, we can see in Fig.~\ref{Fig4} that the \Reimei{} energies converge to energies close to the exact energies. For the rightmost data points shown in Fig.~\ref{Fig4}, we have 
$\delta E_{v_1v_2v_3}^{\text{\Reimei}}<1$ \hcmminusone{} in (a), (d), and (e),
$\delta E_{010}^{\text{\Reimei}}(\Ntrotter = 5) \approx 1.9$ \hcmminusone{} in Fig.~\ref{Fig4}(b), and
$\delta E_{020}^{\text{\Reimei}}(\Ntrotter = 5)  \approx 3.0$ \hcmminusone{} in Fig.~\ref{Fig4}(c). 

Due to the finite number of measurements ($\Nshots=10^3$), the basis sets sampled using \Reimei{} have an inherent statistical variation, meaning that if the simulations were repeated at \Reimei{}, basis sets with a slightly different number of basis functions would be generated. In order to assess the variation of the basis sets due to the finite number of measurements, we have repeated the simulations at the same parameters as used in Fig.~\ref{Fig4} using \Reimei{}-E, a classical emulator of \Reimei{} \cite{RyanAnderson2021,QuantinuumNoiseModel}. The results, presented in Appendix \ref{Appendix:ReimeiEsampling}, show that while there is a slight variation in the number of basis functions, all simulations using \Reimei{}-E result in  data similar to what is  shown in Fig.~\ref{Fig4}.

The energies obtained by sampling the first-order perturbation theory distribution \eqref{Eq:pPTDef}, shown by diamonds in Fig.~\ref{Fig4}, are, at same number of basis functions, 
sometimes lower than the energies obtained from the Suzuki-Trotter distribution \eqref{Eq:pSTDef} [see Fig.~\ref{Fig4}(c) and (e)], 
sometimes comparable [see Fig.~\ref{Fig4}(a) and (b)],
and sometimes higher [see Fig.~\ref{Fig4}(d)]. This suggests that in many cases,  sampling from a distribution determined  by perturbation theory is an efficient classical method to generate compact basis sets which may compete with the QSCI method. However, to obtain eigenenergies that differ by less than 1 \hcmminusone{} from the exact energies, an iterative method  
(like the heat-bath CI method \cite{Holmes2016,Fetherolf2021, Bhatty2021, Tran2023}) where the basis set is systematically enlarged would need to be applied.

As a comparison to the results shown in Fig.~\ref{Fig4}, we have computed the energies using the combined, large basis set as defined in  \eqref{Eq:Omegabig}. We obtain a basis set size of 
$|\Omega_{\text{big}}^{\text{no noise}}|=24$ and $|\Omega_{\text{big}}^{\text{\Reimei}}|=37$ at $\Ntrotter=1$. 
The energies obtained using the large basis sets are summarized in Table.~\ref{Table_bigbasis}. The energy differences $\delta E_{v_1v_2v_3}$ obtained using $\Omega_{\text{big}}^{\text{no noise}}$ are much 
 larger than the noise-free energies shown in Fig.~\ref{Fig4} at $|\Omega|$ around 15.
 In the case of $\Omega_{\text{big}}^{\text{\Reimei}}$, due to the large basis set size, all energy differences except $\delta E_{001}$ are of the same accuracy 
as the \Reimei{} energies in Fig.~\ref{Fig4} at $|\Omega|\approx 20$.

%@@@@@@@@@@@@@@@@@@@@@@@@@@@@@@@@@@@@@@@@@@@@@@@@@@@@@@@@@@@@@@@@@@@@@@@@@@@@@@@@@@@@@@@
\begin{table}
\caption{\label{Table_bigbasis}
Energy differences $\delta E_{v_1v_2v_3}$ in \hcmminusone{} of the combined, large basis set as defined in Eq.~\eqref{Eq:Omegabig}, using $\Ntrotter=1$ and $\vmax=3$. The columns are labeled with the vibrational state $v_1v_2v_3$.
In the lower part of the table, we show for comparison $\delta E_{v_1v_2v_3}$ from Fig.~\ref{Fig4} at $\Ntrotter=5$ for the noise-less result, and $\Ntrotter=4$ for the result obtained at \Reimei{}. The basis set size $|\Omega|$ is also given because $|\Omega|$ is different for each vibrational state. 
}
\begin{ruledtabular}
\begin{tabular}{l|ddddd}
\multicolumn{1}{c|}{} & \multicolumn{1}{c}{$000$} & \multicolumn{1}{c}{$010$} & \multicolumn{1}{c}{$020$} & \multicolumn{1}{c}{$100$} & \multicolumn{1}{c}{$001$}  \\
\hline
No noise (big)                         &        0.7     &   8.0     &   10.3      &      4.8     &    8.5            \\
$|\Omega_{\text{big}}|=24$ &&&&&\\
 Reimei (big)                          &        0.02    &   2.4     &   1.8       &    0.1       &   10.6             \\
 $|\Omega_{\text{big}}|=37$ &&&&&\\
 \hline
No noise (Fig.~\ref{Fig4})             &     0.2        &   1.5     &     4.4    &    1.1   &     1.3            \\
\multicolumn{1}{c|}{$|\Omega|=$} & \multicolumn{1}{c}{14} & \multicolumn{1}{c}{14} & \multicolumn{1}{c}{14} & \multicolumn{1}{c}{15} & \multicolumn{1}{c}{16}  \\
Reimei (Fig.~\ref{Fig4})                       &     0.005    &    2.4    &    3.1    &    1.3   &    1.1             \\
\multicolumn{1}{c|}{$|\Omega|=$} & \multicolumn{1}{c}{24} & \multicolumn{1}{c}{20} & \multicolumn{1}{c}{22} & \multicolumn{1}{c}{20} & \multicolumn{1}{c}{22}  \\
\end{tabular}
\end{ruledtabular}
\end{table}

%@@@@@@@@@@@@@@@@@@@@@@@@@@@@@@@@@@@@@@@@@@@@@@@@@@@@@@@@@@@@@@@@@@@@@@@@@@@@@@@@@@@@@@@

%@@@@@@@@@@@@@@@@@@@@@@@@@@@@@@@@@@@@@@@@@@@@@@@@@@@@@@@@@@@@@@@@@@@@@@@@@@@@@@@@@@@@@@@
\begin{table}
\caption{\label{Table_random}
Average energy difference $\overline{\delta E}_{v_1v_2v_3}$ in \hcmminusone{}  calculated as an average of $\delta E_{v_1v_2v_3}$ obtained from $10^4$ randomly sampled basis sets of varying size $|\Omega_{\rm rand}|=10$, 20 and 30. The statistical uncertainties (one standard deviation) are indicated in parentheses. 
The columns are labeled with the vibrational state $v_1v_2v_3$.}
\begin{ruledtabular}
\begin{tabular}{l|ccccc}
\multicolumn{1}{c|}{} & \multicolumn{1}{c}{$000$} & \multicolumn{1}{c}{$010$} & \multicolumn{1}{c}{$020$} & \multicolumn{1}{c}{$100$} & \multicolumn{1}{c}{$001$}  \\
\hline
$|\Omega_{\rm rand}|=10$    &       21(4)  &  82(17)  &  73(11)  & 430(108)  & 379(93)        \\%4    17    11   108    93
$|\Omega_{\rm rand}|=20$    &       17(6)  &  67(26)  &  60(17)  & 346(167)  & 306(147)         \\%6    26    17   167   147
$|\Omega_{\rm rand}|=30$    &       14(7)  &  53(29)  &  47(19)  & 271(185)  & 241(165)            \\ %7    29    19   185   165
\end{tabular}
\end{ruledtabular}
\end{table}

%@@@@@@@@@@@@@@@@@@@@@@@@@@@@@@@@@@@@@@@@@@@@@@@@@@@@@@@@@@@@@@@@@@@@@@@@@@@@@@@@@@@@@

Some reports \cite{RobledoMoreno2025,Sugisaki2025} compare the energies obtained using  QSCI with the energies obtained from a random basis set sampled from a flat distribution. In our case, the basis sets sampled on \Reimei{} are much better than a random  basis set. 
We define a random basis set of size $|\Omega_{\text{rand}}|$ as the five basis functions $|v_1v_2v_3\rangle$ with $v_1v_2v_3$ in 
$\Gamma$  [see Eq.~\eqref{Eq:initialstatespsi0}] combined with $|\Omega_{\text{rand}}|-5$ randomly selected basis states. 
We assume that the same basis set is used for all five vibrational states in $\Gamma$.
The energy differences averaged over $10^4$ randomly sampled basis sets with sizes $\Omega_{\text{rand}}=10$, 20, and 30 are shown in Table~\ref{Table_random}.
 In all cases, the average energy differences $\overline{\delta E}_{v_1v_2v_3}$ are much larger than the energy differences obtained using \Reimei{} (see Fig.~\ref{Fig4} and Table~\ref{Table_bigbasis}).
  
%...............................................................................................................................................................................................................................................................................................................................................................................................
\subsubsection{
Vibrational energy levels for $\vmax=7$ and $J=0$}
%...............................................................................................................................................................................................................................................................................................................................................................................................

%@@@@@@@@@@@@@@@@@@@@@@@@@@@@@@@@@@@@@@@@@@@@@@@@@@@@@@@@@@@@@@@@@@@@@@@@@@@@@@@@@@@@@@@
\begin{figure}
\includegraphics[width=\figwidth]{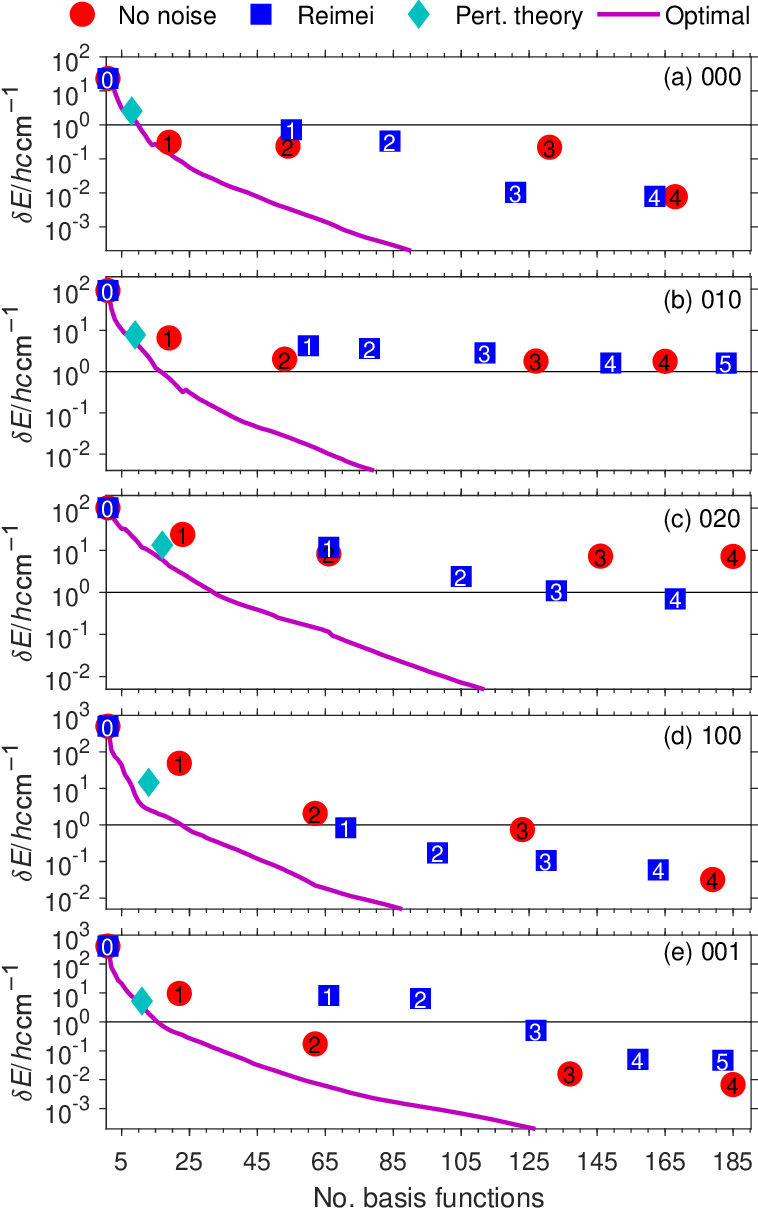}
\caption{\label{Fig5}
Energy differences $\delta E_{v_1v_2v_3}$  on a logarithmic scale for $v_1v_2v_3$  equal to (a) 000, (b) 010, (c) 020, (d) 100, and (e) 001, as a function of the number of basis functions, for $J=0$, $\vmax=7$ ($\nqubit =9$).  The same curve styles  as in Fig.~\ref{Fig4} are employed. 
 The number of Suzuki-Trotter steps is fixed to $\Ntrotter =1$ and the time step $\tau=n\tau_0$ with $\tau_0=0.48$ fs and $n=0, 1, \ldots, 5$ (indicated inside the curve symbols).
}
\end{figure}
%%Data taken on Reimei during Nov 21, 22, Dec 2,3
%@@@@@@@@@@@@@@@@@@@@@@@@@@@@@@@@@@@@@@@@@@@@@@@@@@@@@@@@@@@@@@@@@@@@@@@@@@@@@@@@@@@@@@@

In Fig.~\ref{Fig5}, we show an example of the vibrational energies at $J=0$ obtained with a larger value of the maximum vibrational quantum number $\vmax =7$, requiring $\nqubitvib=3$ qubits per mode and a total of $\nqubit=9$ qubits. The total number of basis states in the full basis set is $(\vmax+1)^3=512$ and the number with correct parity is $512/2=256$. Differently from the calculation shown in Fig.~\ref{Fig4}, here we employ a single Suzuki-Trotter step $(\Ntrotter =1)$ and vary the value of the time step $\tau$. In order to make the simulation feasible at \Reimei{}, the cutoff parameter in Eq.~\eqref{Eq:USTdef} is set to $\lambda=219$ \hcmminusone{}. There are 146 terms in the qubit Hamiltonian  \eqref{Eq:qubitHamiltonian} satisfying $|h_{\bm{k}}|> 219$ \hcmminusone{}: 14 single-qubit,     44 two-qubit,     51 three-qubit,     25 four-qubit, and     12 five-qubit. The simulations using \Reimei{}  were carried out during November 21--22 and December 2--3, 2025. The transpiled circuits contain  around 280 single-qubit gates and 170 two-qubit $R_{ZZ}$ gates. 

We see in Fig.~\ref{Fig5} that the rovibrational QSCI method works rather well also for $\Ntrotter=1$ and a larger value of $\vmax$. The number of sampled basis functions increases with an increasing value of the time step $\tau$, with typically around 20 basis functions sampled at the smallest value of $\tau = 0.48$ fs ($=20\hbar/E_{\rm h}$) in the noise-free case, and about 65 basis functions for \Reimei{}. At the rightmost data points shown in 
Fig.~\ref{Fig5}, we have 
$\delta E_{v_1v_2v_3}^{\text{no noise}}<1$ \hcmminusone{} in (a), (d), and (e),
$\delta E_{010}^{\text{no noise}}(n=4) \approx  1.8$ \hcmminusone{}, and
$\delta E_{020}^{\text{no noise}}(n=4) \approx  7.1$ \hcmminusone{}. 
For the data obtained using \Reimei{}, we have 
$\delta E_{v_1v_2v_3}^{\text{\Reimei}}<1$ \hcmminusone{} in (a), (c), (d), and (e), and
$\delta E_{010}^{\text{\Reimei}}(n=5) \approx  1.6$ \hcmminusone{}. Interestingly, for the $v_1v_2v_3=020$ state 
[see Fig.~\ref{Fig5}(c)], the convergence is better for the basis sets sampled using \Reimei{}, indicating that a certain amount of noise can sometimes be helpful for finding suitable basis functions.

%...............................................................................................................................................................................................................................................................................................................................................................................................
\subsubsection{Rovibrational energy levels for $\vmax=3$ and $J=1$}\label{Subsubec:vmax3J1}
%...............................................................................................................................................................................................................................................................................................................................................................................................
In this section, we consider the case of a non-zero value of the total angular momentum $J$. For the low-energy eigenvalues $\Delta E<4000$ \hcmminusone{} we are considering here, the rovibrational eigenstates $|\phi\rangle$ can be approximated as a product of a vibrational wave function and an asymmetric top rotational wave function $|\Phi_{J_{K_a K_c}}\rangle$,
\begin{equation}\label{Eq:vibrotPhiapprox}
|\phi\rangle \approx |\phi_{\text{vib}}\rangle |\Phi_{J_{K_a K_c}}\rangle.
\end{equation}
 The asymmetric top wave function $|\Phi_{J_{K_a K_c}}\rangle$ is an eigenfunction of the rigid-rotor Hamiltonian $H_{\text{RR}}$ [see Eq.~\eqref{Eq:HRRDef}], and can be expressed as a superposition of the angular momentum basis functions $|J,K\rangle$. For example, for $J=1$, with our definition of the $c$ axis as the $C_2$ axis, we have 
 \begin{align}\label{Eq:AsymRotPhiJ1}
 |\Phi_{1_{01}}\rangle&=\frac{1}{\sqrt{2}}(|1,1\rangle +|1,-1\rangle ),
 \nonumber \\
 |\Phi_{1_{11}}\rangle&=|1,0\rangle, \text{ and}
 \nonumber \\
 |\Phi_{1_{10}}\rangle&=\frac{1}{\sqrt{2}}(|1,1\rangle -|1,-1\rangle ).
 \end{align}
The labels $J_{K_aK_c}$ on $|\Phi_{J_{K_aK_c}}\rangle$ are defined by the quantum numbers of the prolate and oblate symmetric top limits (when two rotational constants coincide) \cite{Bauder2011}. In order of lowest to highest energy, the order of the labels is $J_{0,J}$, $J_{1,J}$, $J_{1,J-1}$, $J_{2,J-1}$, $\ldots$, $J_{J-1,1}$, $J_{J,1}$, $J_{J,0}$ \cite{Bauder2011}.

Because of Eq.~\eqref{Eq:vibrotPhiapprox}, it is reasonable  to employ the same  basis sets as sampled for $J=0$ also for $J>0$. The basis sets for $J>0$ are taken to be
\begin{equation}
\Omega_{v_1v_2v_3,J}=\{|v'_1v'_2v'_3\rangle |J,K\rangle \},
\end{equation}
with
\begin{equation}
|v'_1v'_2v'_3\rangle \in \Omega_{v_1v_2v_3}, \text{ and } K=-J,\ldots,J, 
\end{equation}
where  $\Omega_{v_1v_2v_3}$ is a basis set sampled for $J=0$ as described in Sec.~\ref{Subsec:QSCI}. The size of the basis set is $|\Omega_{v_1v_2v_3,J}|=|\Omega_{v_1v_2v_3}|(2J+1)$.
The eigenenergies are obtained similarly as for $J=0$, that is, we first obtain $2J+1$ eigenstates for each basis set $\Omega_{v_1v_2v_3,J}$, and then combine the eigenstates from all five basis sets into a final basis set which is orthogonalized. The final estimates of the energies are obtained by diagonalizing the $5(2J+1)\times 5(2J+1)$ Hamiltonian matrix. A rovibrational eigenstate $|\phi\rangle$ is labeled by $v_1v_2v_3J_{K_aK_c}$,  determined by the state $|v_1v_2v_3\rangle |\Phi_{J_{K_aK_c}}\rangle$ that has the largest overlap with $|\phi\rangle$.

Although not considered here, we mention that for $J>0$, the size of the Hamiltonian matrix can be reduced by 
employing basis functions which are symmetric or antisymmetric under  spatial inversion. The inversion symmetry could also be useful for post-selection of basis states in a case where a rovibrational trial wave function with $J>0$ is employed for the basis set sampling.

%@@@@@@@@@@@@@@@@@@@@@@@@@@@@@@@@@@@@@@@@@@@@@@@@@@@@@@@@@@@@@@@@@@@@@@@@@@@@@@@@@@@@@@@
\begin{figure}
\includegraphics[width=\figwidth]{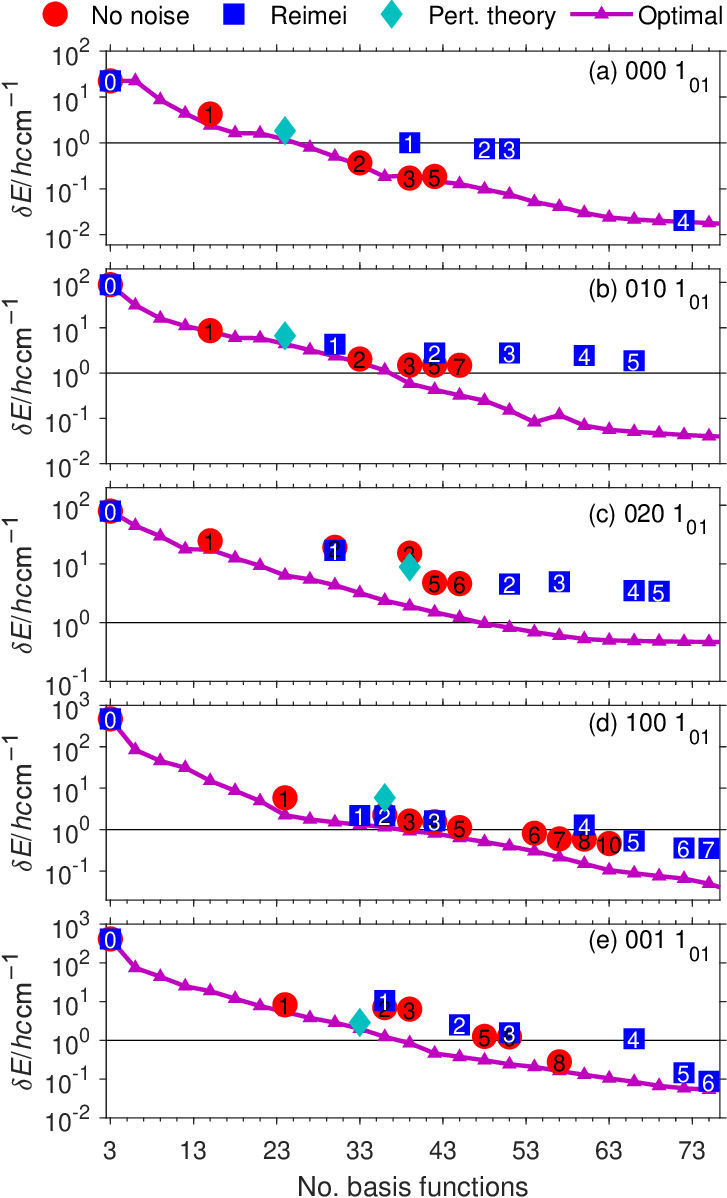}
\caption{\label{Fig6}
Rovibrational energy differences $\delta E_{v_1v_2v_3J_{K_aK_c}}$ for $J=1$, and otherwise the same parameters as in Fig.~\ref{Fig4} ($\vmax=3$, fixed $\tau=0.24$ fs, varying $\Ntrotter$ as indicated inside the curve symbols).  The state label $v_1v_2v_3J_{K_aK_c}$ is indicated in the upper right corner inside each panel. Only the $J_{K_aK_c}=1_{01}$ states are shown.
}
\end{figure}
%@@@@@@@@@@@@@@@@@@@@@@@@@@@@@@@@@@@@@@@@@@@@@@@@@@@@@@@@@@@@@@@@@@@@@@@@@@@@@@@@@@@@@@@

In Fig.~\ref{Fig6}, we show an example  for $J=1$, $\vmax=3$, $\tau=0.24$ fs, and $\Ntrotter=0,1,\ldots,10$ (the same parameters as in Fig.~\ref{Fig4} except for $J$). The total number of basis functions is $(2J+1)(\vmax+1)^3= 192$. Because $J=1$, there are now $2J+1=3$ rovibrational states labeled with the same vibrational quantum numbers $v_1v_2v_3$.  

We find that a similar accuracy $\delta E_{v_1v_2v_3J_{K_aK_c}}$ is achieved for all three rotational states, and we therefore show only the lowest-energy rotational state ($J_{K_aK_c}=1_{01}$) in Fig.~\ref{Fig6}. Here $\delta E_{v_1v_2v_3J_{K_aK_c}}$ is defined  as
\begin{equation}
    \delta E_{v_1v_2v_3J_{K_aK_c}} = E_{v_1v_2v_3J_{K_aK_c}}-E_{v_1v_2v_3J_{K_aK_c}}^{\rm exact},
\end{equation}
 similarly to Eq.~\eqref{Eq:DefDeltaE}. A summary of the rovibrational energies and energy differences for all
 rovibrational states is given in Table~\ref{Table:rovibstates}.

%@@@@@@@@@@@@@@@@@@@@@@@@@@@@@@@@@@@@@@@@@@@@@@@@@@@@@@@@@@@@@@@@@@@@@@@@@@@@@@@@@@@@@@@
\begin{table}
\caption{\label{Table:rovibstates}
Upper part: Exact rovibrational energies $\Delta E_{v_1v_2v_3J_{K_aK_c}}=E_{v_1v_2v_3J_{K_aK_c}}-E_{000\,0_{00}}$ in \hcmminusone{}, defined relative to the rovibrational ground state. Lower part: Energy differences $\delta E_{v_1v_2v_3J_{K_aK_c}}$ in \hcmminusone{}
for basis sets sampled without noise, and with Reimei. The data for $J_{K_aK_c}=1_{01}$ is the same as shown in Fig.~\ref{Fig6} (rightmost symbols).
The basis set sizes $|\Omega|$ are indicated for each vibrational state.
}
\begin{ruledtabular}
\begin{tabular}{c|ddddd}
\multicolumn{1}{c|}{} & \multicolumn{1}{c}{$000$} & \multicolumn{1}{c}{$010$} & \multicolumn{1}{c}{$020$} & \multicolumn{1}{c}{$100$} & \multicolumn{1}{c}{$001$}  \\
\hline
$\Delta E$ (exact) &&&&&\\
$1_{01}$ & 25.3 & 1615.3 &  3186.9 & 3742.3 & 3823.8 \\
$1_{11}$ & 39.3 &  1632.0 & 3206.1 & 3756.7 & 3837.5 \\
$1_{10}$ & 44.9 &  1638.0 & 3212.3 & 3762.7 & 3843.6 \\
\hline
$\delta E$ (no noise) &&&&&\\
$1_{01}$ & 0.2 & 1.5 & 4.6 & 0.5 & 0.3 \\
$1_{11}$ & 0.2 & 1.5 &  4.7 & 0.5 & 0.3  \\
$1_{10}$ & 0.2 &  1.5 & 4.2  & 0.5 & 0.3 \\
%42 45 45 63 57
\multicolumn{1}{c|}{$|\Omega|=$} & \multicolumn{1}{c}{42} & \multicolumn{1}{c}{45} & \multicolumn{1}{c}{45} & \multicolumn{1}{c}{63} & \multicolumn{1}{c}{57}  \\
$\delta E$ (Reimei) &&&&&\\
$1_{01}$ & 0.02 & 1.9 & 3.4 & 0.4 & 0.09 \\
$1_{11}$ & 0.02 & 1.9 & 3.3 & 0.4 & 0.1 \\
$1_{10}$ & 0.006 & 1.9 &  2.9 & 0.3 & 0.04\\
%72 66 69 75 75 
\multicolumn{1}{c|}{$|\Omega|=$} & \multicolumn{1}{c}{72} & \multicolumn{1}{c}{66} & \multicolumn{1}{c}{69} & \multicolumn{1}{c}{75} & \multicolumn{1}{c}{75}  \\
\end{tabular}
\end{ruledtabular}
\end{table}

%@@@@@@@@@@@@@@@@@@@@@@@@@@@@@@@@@@@@@@@@@@@@@@@@@@@@@@@@@@@@@@@@@@@@@@@@@@@@@@@@@@@@@@@

Naturally, a similar energy convergence with an increasing  number of basis functions as in Fig.~\ref{Fig4}  is observed in Fig.~\ref{Fig6}. For the rightmost data points in Fig.~\ref{Fig6}, we have 
$\delta E_{v_1v_2v_31_{01}}<1$ \hcmminusone{} in  Fig.~\ref{Fig6}(a), (d), and (e),
$\delta E^{\text{no noise}}_{010\,1_{01}}\approx 1.5$ \hcmminusone{} and 
$\delta E^{\text{\Reimei}}_{010\,1_{01}}\approx 1.9$ \hcmminusone{}
in  Fig.~\ref{Fig6}(b), and
$\delta E^{\text{no noise}}_{020\,1_{01}}\approx 4$ \hcmminusone{} and 
$\delta E^{\text{\Reimei}}_{020\,1_{01}}\approx 3$ \hcmminusone{}
in  Fig.~\ref{Fig6}(c).

The proton exchange parity can be expressed in terms of  $v_3$, $K_a$ and $K_c$ as $P_{\rm ex}=(-1)^{v_3+K_a+K_c}$ [in view of \eqref{Eq:AsymRotPhiJ1}, this is consistent with \eqref{Eq:ProtonExParityDef}]. Therefore, for $v_1v_2v_3=000$, 010, 020, and 100, the $J_{K_aK_c}=1_{10}$ and $1_{01}$  states have proton spin $s=1$, and the $1_{11}$ has $s=0$, while for $v_1v_2v_3=001$,   $1_{10}$ and $1_{01}$ are singlet states and $1_{11}$ is a triplet state.

The results shown in Fig.~\ref{Fig6} demonstrate that the Hamiltonian-simulation based  rovibrational QSCI method can be used to generate rather compact basis sets for rovibrational states also at non-zero values of the total angular momentum $J$.

%=========================================================================================================================
\section{Summary and outlook}
%=========================================================================================================================
We have described how to calculate approximate eigenstates of  rotating and vibrating molecules using quantum computers. The transformation of the Watson Hamiltonian into qubit form revealed that the qubit Hamiltonian expressed as a linear combination of Pauli strings contains thousands of terms, too many to apply standard quantum computing methods like the VQE. By adopting a variant of the QSCI method, where the basis set is sampled on a quantum computer and the energies are computed classically, we showed that rovibrational energy levels of H$_2$O can be evaluated to an accuracy of below 1 \hcmminusone{} in many cases. 
The main advantage of the QSCI method over the VQE method is that the energy levels can be estimated using relatively few circuit measurements, and that the construction and diagonalization of the Hamiltonian matrix does not suffer from errors due to the noise. Another advantage is that, as shown in Sec.~\ref{Subsubec:vmax3J1}, the treatment of rotating molecules ($J>0$) is straightforward.

While our results suggest that current noisy quantum computers may be of use in rovibrational structure calculations, there are many directions left for future investigations. To assess the true strength of  the rovibrational QSCI method, we would need to consider large molecules (e.g.\ benzene \cite{Halverson2015,Wang_SciRep2020}) or molecular clusters (such as water clusters \cite{Yagi2017,Zhang2023,Simko2025}) requiring large basis sets so that the Hamiltonian matrix becomes too large to be exactly diagonalized. 
We comment that the circuit depth ($\sim 200$ two-qubit gates) considered in the present investigation is far from the limits of what can be executed on existing trapped-ion quantum computers. For example, in a recent investigation using Quantinuum's 56-qubit H2 device, circuits containing more than 2000 two-qubit gates were executed \cite{Alam2025}. For the sampling of larger basis sets, a larger number of shots $\Nshots$ will be necessary. On Reimei, it is practically difficult to increase the number of shots to  values larger than $\Nshots>10^4$. Further research on QSCI applied to the rovibrational structure problem will need to address the problem of a limited number of shots for sampling large basis sets.

The most efficient qubit mapping and basis set also deserve further attention \cite{Mikkelsen2025}. We have used a harmonic oscillator basis in this paper, but it is possible that local, grid-type basis sets \cite{Hanada2025,Szczepanik2025} 
or an $n$-body expansion of the potential \cite{Kongsted2006,Ollitraultetal2020} combined with an representation of the vibrational wave function using modals \cite{Christiansen2004,Ollitraultetal2020} 
can be more efficient  for current noisy quantum computers.

Rovibrational structure simulations have an advantage over other types of    simulations because we can  compare the  results of the simulations with high-precision experimental results in a direct way. This suggests that once rovibrational structure calculations can be reliably carried out on quantum computers, they can be suitable for a verifiable quantum advantage in chemistry:
A quantum computer is used to calculate rovibrational transitions in a molecule which is too large to treat classically, but for which transition frequencies can be measured by spectroscopic methods to high precision.

%=========================================================================================================================
\acknowledgments
We thank Xiaoyang Wang and Arata Yamamoto (RIKEN iTHEMS and RIKEN Quantum) for useful comments. We thank Tomoya Okino (RIKEN) and Nathan Lysne (Quantinuum) for providing the data in Appendix \ref{Appendix:ReimeiParameters}.
E.~L.\ is supported by the JSPS (Kakenhi no.~JP24K08336), by NEDO (project JPNP20017), and by the RIKEN TRIP initiative (RIKEN Quantum).
T.~Sz.\ is supported by the NKFIH grant NKKP ADVANCED 152731.
%=========================================================================================================================

%=========================================================================================================================
\section{Author declarations}
\subsection{Conflict of interest}
The authors have no conflicts to disclose.
%=========================================================================================================================

%=========================================================================================================================
\section{Data availability}
The data supporting the findings of this article are openly
available (see Ref.~\onlinecite{Lotstedt_RovibDataset}).
%=========================================================================================================================

\appendix

%=========================================================================================================================
\section{Definitions of terms in the Watson Hamiltonian}\label{Appendix:Defs}
%=========================================================================================================================
We largely follow Ref.~\onlinecite{Bauder2011}, although we adopt a slightly different notation for the angular momentum operators.
We fix a coordinate system $\bm{r}=(r_a,r_b, r_c)$ in the molecular frame, denote the coordinate of atom $n$ in the equilibrium geometry of the molecule by $\bm{r}_n^{\rm e}$, and an atomic coordinate displaced from the equilibrium  by 
\begin{equation}
\bm{r}_n=\bm{r}_n^{\rm e}+\bm{\rho}_n,
\end{equation}
where $\bm{r}_n^{\rm e}$ and $\bm{\rho}_n$ satisfy the Eckart conditions \cite{Eckart1935} $\sum_nm_n\bm{\rho}_n=\bm{0}$ and 
$\sum_nm_n\bm{r}_n^{\rm e}\times \bm{\rho}_n=\bm{0}$ ($m_n$ is the mass of atom $n$).
  The mass-scaled normal mode coordinates $Q_k$ are defined  as
\begin{equation}\label{Eq:Qkdef}
Q_k = \sum_{n=1}^{\Natoms}\sum_{\alpha=a,b,c}L_{n\alpha,k}\sqrt{m_n}\rho_{n\alpha},
\end{equation}
where $L_{n\alpha,k}$ is the transformation matrix determined by the condition that the potential energy $V(Q_1,\ldots,Q_{\Nmodes})$ only contains terms quadratic in $Q_k$ for small $Q_k$.

The Coriolis coupling coefficients are given by 
\begin{equation}\label{Eq:CorioloiscouplDef}
\zeta_{kl}^\alpha=\sum_{n=1}^{\Natoms}\sum_{\beta,\gamma=a,b,c} \varepsilon_{\alpha\beta\gamma}L_{n\beta,k}L_{n\gamma,l},
\end{equation}
where $\varepsilon_{\alpha\beta\gamma}$ is the Levi-Civita symbol. The $\zeta_{kl}^\alpha$ matrices are antisymmetric, $\zeta_{kl}^\alpha=-\zeta_{lk}^\alpha$.

 The inverse inertia tensor $\bm{\mu}$ is defined as the inverse 
of the tensor $\bm{I}'$,
\begin{equation}
\bm{\mu}={\bm{I}'}^{-1},
\end{equation}
where ${\bm{I}'}$ has the molecular-frame matrix elements
\begin{equation}\label{Eq:defIprime}
I'_{\alpha\beta}=I_{\alpha\beta}-\sum_{k,l,m=1}^{\Nmodes}\zeta_{km}^\alpha\zeta_{lm}^\beta Q_k Q_l.
\end{equation}
In Eq.~\ref{Eq:defIprime}, $I_{\alpha\beta}$ is the usual inertia tensor defined using the atomic positions as
\begin{equation}
I_{\alpha\beta}=\sum_{n=1}^{\Natoms}m_n \left(\bm{r}_n^2 \delta_{\alpha\beta} - r_{n\alpha}r_{n\beta}\right).
\end{equation}
The inertia tensor at the equilibrium geometry is denoted by $\bm{I}^{\rm e}$ and is defined similarly,
\begin{equation}
I^{\rm e}_{\alpha\beta}=\sum_{n=1}^{\Natoms}m_n \left({\bm{r}_n^{\rm e}}^2 \delta_{\alpha\beta} - r_{n\alpha}^{\rm e}r_{n\beta}^{\rm e}\right).
\end{equation}
The molecular-frame coordinate system is selected so that $\bm{I}^{\rm e}$ is diagonal, that is $I^{\rm e}_{\alpha\beta}=0$ for $\alpha\ne\beta$. The equilibrium rotational constants for rotation around axis $\alpha$ is given by  $\hbar^2/(2I^{\rm e}_{\alpha\alpha})$. 

The inverse inertia tensor $\bm{\mu}$ can be expanded in orders of $Q_k$ as indicated in Eq.~\eqref{Eq:muexpansion}. The different orders are given by 
\begin{equation}\label{Eq:muorders}
\bm{\mu}_\ell = c_\ell {\bm{I}^{\rm e}}^{-\frac{1}{2}} \bm{b}^\ell {\bm{I}^{\rm e}}^{-\frac{1}{2}},
\end{equation}
where the coefficients $c_\ell$ are defined as 
\begin{equation}
c_0=1,\quad c_1 =-1,\quad c_2 = \frac{3}{4}, \quad c_3 = -\frac{1}{2}, \quad c_4 = \frac{5}{16},
\end{equation}
\begin{equation}
\bm{b}=\sum_{k=1}^{\Nmodes} {\bm{I}^{\rm e}}^{-\frac{1}{2}} \bm{a}_k {\bm{I}^{\rm e}}^{-\frac{1}{2}} Q_k,
\end{equation}
and the matrix elements of $\bm{a}_k$ are defined as
\begin{equation}\label{Eq:akDef}
a_{k\,\alpha\beta} = 2\sum_{n=1}^{\Natoms}\sum_{\gamma,\delta=a,b,c}(\delta_{\alpha\beta}\delta_{\gamma\delta}-\delta_{\alpha\delta}\delta_{\beta\gamma})
\sqrt{m_n}r^{\rm e}_{n\gamma} L_{n\delta,k}.
\end{equation}

The angular momentum operators $J_{a,b,c}$ in the molecular frame  satisfy 
\begin{equation}\label{Eq:Jcommutator}
[J_\alpha,J_\beta]=-i\hbar J_\gamma, 
\end{equation}
where $(\alpha,\beta,\gamma)$ is a cyclic permutation of $(a,b,c)$. The angular momentum states $|J,K\rangle$ are eigenstates of $J_c$ and $\bm{J}^2=J_a^2+J_b^2+J_c^2$,
\begin{align}
J_c|J,K\rangle&=\hbar K|J,K\rangle, \nonumber \\
\bm{J}^2|J,K\rangle&=\hbar^2J(J+1)|J,K\rangle.
\end{align}
Differently from the lab frame, Eq.~\eqref{Eq:Jcommutator} has a minus sign on the right-hand side. This means that for the raising and lowering operators $J^\pm = J_a\pm iJ_b$, we have 
\begin{equation}
J^{\pm}|J,K\rangle=\hbar\sqrt{J(J+1)-K(K\mp 1)}|J,K\mp 1\rangle,
\end{equation}
that is, $J^+$ lowers, and $J^-$ raises the quantum number $K$. 

%=========================================================================================================================
\section{Parameters of H$_2$O}\label{Appendix:H2Oparameters}
%=========================================================================================================================
We employ the following structural parameters of H$_2$O, obtained at the  CCSD(T)/aug-cc-pVQZ level using {\sc Molpro} \cite{MOLPRO,Werner2011,Werner2020}.

Equilibrium geometry: O--H internuclear distance 
\begin{equation}
|\bm{r}^{\rm e}_{\rm O} - \bm{r}^{\rm e}_{\rm H}|=1.8121 a_0,
\end{equation}
where $a_0\approx 0.529$ \AA{} is the Bohr radius,
 H--O--H bond angle 
 \begin{equation}
 \theta_{\rm OHO} = 104.368^\circ.
 \end{equation}

Equilibrium rotational constants:
\begin{align}
A^{\rm e}&=\frac{\hbar^2}{2I^{\rm e}_{aa}} = 9.49 \text{ \hcmminusone{}},
\nonumber \\
B^{\rm e}&=\frac{\hbar^2}{2I^{\rm e}_{bb}} = 27.24  \text{ \hcmminusone{}, and}
\nonumber \\
C^{\rm e}&=\frac{\hbar^2}{2I^{\rm e}_{cc}} = 14.57 \text{ \hcmminusone{}}.      
\end{align}
 Note that  our definition of the $a$, $b$ and $c$ axes is different from the usual convention for asymmetric top molecules, in which the $a$, $b$ and $c$ axes are selected so that 
 $A^{\rm e}>B^{\rm e}>C^{\rm e}$ \cite{Bauder2011}.
 
Harmonic frequencies (symmetric stretch, bending, antisymmetric stretch): 
\begin{equation}
\hbar(\omega_1,\omega_2,\omega_3)= (3830.87,1649.74, 3940.46) \text{ \hcmminusone{}}.
\end{equation}

Anharmonic coupling constants (only non-zero values of terms $g_{jkl}$ and $f_{jklm}$ with $j\le k \le l \le m$ are given):
\begin{align}\label{Eq:gijk}
g_{133}/(\hbar^{-\frac{3}{2}}\omega_1^{\frac{1}{2}}\omega_3)&=-1822.17 \text{ \hcmminusone{}}, \\\nonumber
g_{233}/(\hbar^{-\frac{3}{2}}\omega_2^{\frac{1}{2}}\omega_3)&=-265.63 \text{ \hcmminusone{}}, \\\nonumber
g_{111}/(\hbar^{-\frac{3}{2}}\omega_1^{\frac{3}{2}})&=-1843.08 \text{ \hcmminusone{}}, \\\nonumber
g_{112}/(\hbar^{-\frac{3}{2}}\omega_1\omega_2^{\frac{3}{2}})&=-73.77  \text{ \hcmminusone{}}, \\\nonumber
g_{122}/(\hbar^{-\frac{3}{2}}\omega_1^{\frac{1}{2}}\omega_2)&=310.01 \text{ \hcmminusone{}}, \\\nonumber
g_{222}/(\hbar^{-\frac{3}{2}}\omega_2^{\frac{3}{2}})&=264.59 \text{ \hcmminusone{}},
\end{align}
\begin{align}\label{Eq:fijkl}
f_{3333}/(\hbar^{-2}\omega_3^2)&=779.10  \text{ \hcmminusone{}}, \\\nonumber
f_{1133}/(\hbar^{-2}\omega_1\omega_3)&=764.90 \text{ \hcmminusone{}}, \\\nonumber
f_{1111}/(\hbar^{-2}\omega_1^2)&=767.10 \text{ \hcmminusone{}}, \\\nonumber
f_{1233}/(\hbar^{-2}\omega_1^{\frac{1}{2}}\omega_2^{\frac{1}{2}}\omega_3)&=118.20 \text{ \hcmminusone{}}, \\\nonumber
f_{1112}/(\hbar^{-2}\omega_1^{\frac{3}{2}}\omega_2^{\frac{1}{2}})&=62.20 \text{ \hcmminusone{}}, \\\nonumber
f_{2233}/(\hbar^{-2}\omega_2\omega_3)&=-368.56 \text{ \hcmminusone{}}, \\\nonumber
f_{1122}/(\hbar^{-2}\omega_1\omega_2)&=-305.41 \text{ \hcmminusone{}}, \\\nonumber
f_{1222}/(\hbar^{-2}\omega_1^{\frac{1}{2}}\omega_2^{\frac{3}{2}})&=-154.64 \text{ \hcmminusone{}}, \\\nonumber
f_{2222}/(\hbar^{-2}\omega_2^2)&=-52.89 \text{ \hcmminusone{}}.
\end{align}
Note that we have to account for the multiplicity of the equivalent terms when using the above numerical values in Eq.~\eqref{Eq:VanharmDef}.
 Except for $f_{2222}$, the numerical values given in Eqs.~\eqref{Eq:gijk} and \eqref{Eq:fijkl} agree with the values in table 6 in  Ref.~\onlinecite{Csaszar1997} to within 20 \hcmminusone{}. The value of $f_{2222}$ is given in  Ref.~\onlinecite{Csaszar1997} as $\hbar^2 f_{2222}/\omega_2^2\approx -11$ \hcmminusone{}. The signs of the coupling constants in  Eqs.~\eqref{Eq:gijk} and \eqref{Eq:fijkl} and table 6 in  Ref.~\onlinecite{Csaszar1997} agree if we redefine the normal mode coordinates as $Q_1\to-Q_1$ and $Q_2\to -Q_2$.

Normal mode transformation matrix [used in Eqs.~\eqref{Eq:Qkdef}, \eqref {Eq:CorioloiscouplDef}, and \eqref{Eq:akDef}]:
\begin{equation}
\bm{L}=
\begin{pmatrix}
\bm{L}_{\rm O} \\\bm{L}_{\rm H1}\\\bm{L}_{\rm H2}
\end{pmatrix},
\end{equation}
where
\begin{align}
\bm{L}_{\rm O}&=
\begin{pmatrix}
L_{\text{O} a,1} & L_{\text{O} a,2}& L_{\text{O} a,3}\\
L_{\text{O} b,1} & L_{\text{O} b,2}& L_{\text{O} b,3}\\
L_{\text{O} c,1} & L_{\text{O} c,2}& L_{\text{O} c,3}\\
\end{pmatrix} \\\nonumber
&=
\begin{pmatrix}
     0          &     0            &      0 \\
     0          &     0            &    -0.2700\\
    0.1949      &    -0.2719      &        0
\end{pmatrix},
\end{align}
\begin{equation}
\bm{L}_{\rm H1}=
\begin{pmatrix}
     0         &    0           &   0 \\
     0.5747 &  0.4120  &  0.5379 \\
     -0.3883 & 0.5416   & -0.4174 
\end{pmatrix},
\end{equation}
and
\begin{equation}
\bm{L}_{\rm H2}=
\begin{pmatrix}
     0         &    0           &   0 \\
     -0.5747  &  -0.4120  &  0.5379 \\
    -0.3883 &  0.5416  &  0.4174
\end{pmatrix}.
\end{equation}
$\bm{L}$ is dimensionless and satisfies $\bm{L}^T\bm{L} =\bm{1}$.
%Orthogonal, dimensionless L matrix, from MATLAB output.
%We are using the low-to-high energy mode ordering below 1: bend, 2: sym strecth, 3: asym stretch
%In the manuscript, we have standard mode ordering 1: sym stretch , 2: bend, 3: asym stretch
%L =
%
 %        0         0         0
 %       0         0   -0.2700
 % -0.2719    0.1949         0
 %      0         0         0
 % 0.4120    0.5747    0.5379
 % 0.5416   -0.3883   -0.4174
 %     0         0         0
 %-0.4120   -0.5747    0.5379
 % 0.5416   -0.3883    0.4174
    
%=========================================================================================================================
\section{$\vmax$-dependence of the number of terms in the qubit Hamiltonian}\label{Appendix:vmaxscaling}
%=========================================================================================================================
%@@@@@@@@@@@@@@@@@@@@@@@@@@@@@@@@@@@@@@@@@@@@@@@@@@@@@@@@@@@@@@@@@@@@@@@@@@@@@@@@@@@@@@@
\begin{figure}
\includegraphics[width=\figwidth]{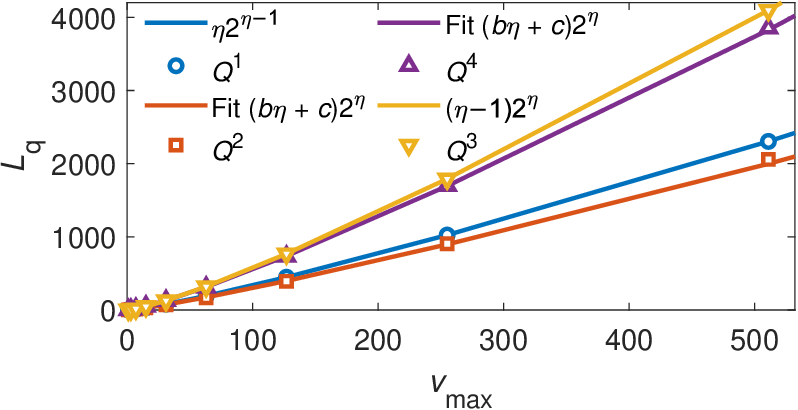}
\caption{\label{Fig1_Appendix}
Open symbols: the number of terms $\NHq$ in the qubit representation of the operator $Q^\ell$ for $\ell=1$, 2, 3, and 4, as a function of the maximum vibrational quantum number $\vmax=2^{\vmaxexponent}-1$ ($\vmaxexponent=1,\ldots,9$). Solid lines: analytic formulas $\NHq=\vmaxexponent 2^{\vmaxexponent-1}$ for $\ell=1$, $\NHq=(\vmaxexponent-1)2^{\vmaxexponent}$ for $\ell=3$, and fits to $\NHq=(b\vmaxexponent+c)2^{\vmaxexponent}$ for $\ell=2$ and 4. We obtain $(b,c)_{\ell=2} = (0.402, 0.288)$ and $(b,c)_{\ell=4} = (0.857, -0.239)$.
}
\end{figure}
%@@@@@@@@@@@@@@@@@@@@@@@@@@@@@@@@@@@@@@@@@@@@@@@@@@@@@@@@@@@@@@@@@@@@@@@@@@@@@@@@@@@@@@@

In this section, we investigate the number of terms in the qubit representation of the matrices $\bm{Q}^{(\ell)}$ having matrix elements
\begin{equation}
(\bm{Q}^{(\ell)})_{vv'} =\langle v|Q^\ell |v'\rangle,
\end{equation}
where $|v\rangle$ is a harmonic-oscillator basis function, and $Q$ is a normal-mode coordinate. In Fig.~\ref{Fig1_Appendix}, we show the number of terms $\NHq$ in the qubit representation of $\bm{Q}^{(\ell)}$,
\begin{equation}
\bm{Q}^{(\ell)} = \sum_{\bm{k}} q^{(\ell)}_{\bm {k}} P_{\bm{k}},
\end{equation}
where $q^{(\ell)}_{\bm {k}}$ is a numerical coefficient, and $P_{\bm{k}}$ is a direct product of Pauli matrices as defined in Eq.~\eqref{Eq:PaulistringDef}. We show $\NHq$ as a function of the maximum value $\vmax$ of the vibrational quantum number, evaluated at values of $\vmax$ such that $\vmax+1=2^\vmaxexponent$ for $\vmaxexponent=1,2,\ldots ,9$. In this case, all qubit states represent harmonic oscillator states in the binary mapping.

In Ref.~\onlinecite{Hanada2025}, it was shown that $\NHq(\bm{Q}^{(1)})=\vmaxexponent2^{\vmaxexponent-1}$. We find that $\NHq(\bm{Q}^{(3)})=(\vmaxexponent-1)2^{\vmaxexponent}$ is valid for $\vmaxexponent>1$. We could not find any analytic formula for $\NHq(\bm{Q}^{(2)})$ and $\NHq(\bm{Q}^{(4)})$. As shown in Fig.~\ref{Fig1_Appendix}, $\NHq(\bm{Q}^{(\ell)})$ ($\ell=2,4$) can be fitted rather well with the expression $\NHq=(b\vmaxexponent+c)2^{\vmaxexponent}$. The values of $b$ and $c$ are obtained by fitting the numerically evaluated values of $\NHq/2^\vmaxexponent$ for $\vmaxexponent>1$ to a linear function of $\vmaxexponent$.

Based on the results shown in Fig.~\ref{Fig1_Appendix}, we conclude that at least up to $\vmax<500$, $\NHq$ scales with $\vmax$ approximately as $\NHq=\bigO[\vmax\log_2(\vmax)]$. 
For operators involving several modes of the kind $Q_1^{\ell_1}Q_2^{\ell_2}Q_3^{\ell_3}$, we have the scaling $\NHq=\bigO[\vmax^{3}\log_2^{3}(\vmax)]$. 

%=========================================================================================================================
\section{Error rates of \Reimei{}}\label{Appendix:ReimeiParameters}
%=========================================================================================================================
In Table~\ref{Table2}, we show the error rates of \Reimei{} at the time of the simulations presented in the paper. The dominant error is the two-qubit gate error, which is of the order of $10^{-3}$ for all simulations.
For a more detailed discussion of the error characterization of Quantinuum's quantum computers, see Refs.~\onlinecite{QuantinuumReimei2025,Moses2023_PRX}. 

%@@@@@@@@@@@@@@@@@@@@@@@@@@@@@@@@@@@@@@@@@@@@@@@@@@@@@@@@@@@@@@@@@@@@@@@@@@@@@@@@@@@@@@@
\setlength{\tabcolsep}{-2pt}
\begin{table}[h]
\caption{\label{Table2} Error rates of \Reimei{} at different execution dates: I (July 18--26, 2025), II (November 21--22, 2025), 
and III (December 2--3, 2025). The value in parentheses indicates the experimental uncertainty. }
\begin{ruledtabular}
\begin{tabular}{lddd}
                                 &  \multicolumn{1}{c}{I}     & \multicolumn{1}{c}{II}       & \multicolumn{1}{c}{III}              \\
\hline
1-qubit gate error$/10^{-5}$     &  1.8(3)      &    2.3(3)       &  2.8(5)                   \\
1-qubit leakage error$/10^{-6}$  &  2\rlap{(1)} &    4\rlap{(1)}  &  3\rlap{(1)}               \\
2-qubit gate error$/10^{-3}$     &  1.35(4)     &    1.22(5)       &  1.30(1)                  \\
2-qubit leakage error$/10^{-4}$  &  1.9(2)      &    1.6(2)       &  1.9(1)      \\
Memory error$/10^{-4}$           & 2.4(6)       &  1.3(2)         &  1.5(2)                    \\
Meas. crosstalk error$/10^{-5}$  &  1.00(7)     &   0.5(1)        &   0.34(4)                   \\
SPAM\footnote{State preparation and measurement} error (av.)$/10^{-3}$   
                                 & 3.7(2)       &  4.3(1)         &  3.7(2)                    \\
\end{tabular}
\end{ruledtabular}
\end{table}
%@@@@@@@@@@@@@@@@@@@@@@@@@@@@@@@@@@@@@@@@@@@@@@@@@@@@@@@@@@@@@@@@@@@@@@@@@@@@@@@@@@@@@@@

%=========================================================================================================================
\section{Sampling using \Reimei{}-E}\label{Appendix:ReimeiEsampling}
\begin{figure}
\includegraphics[width=\figwidth]{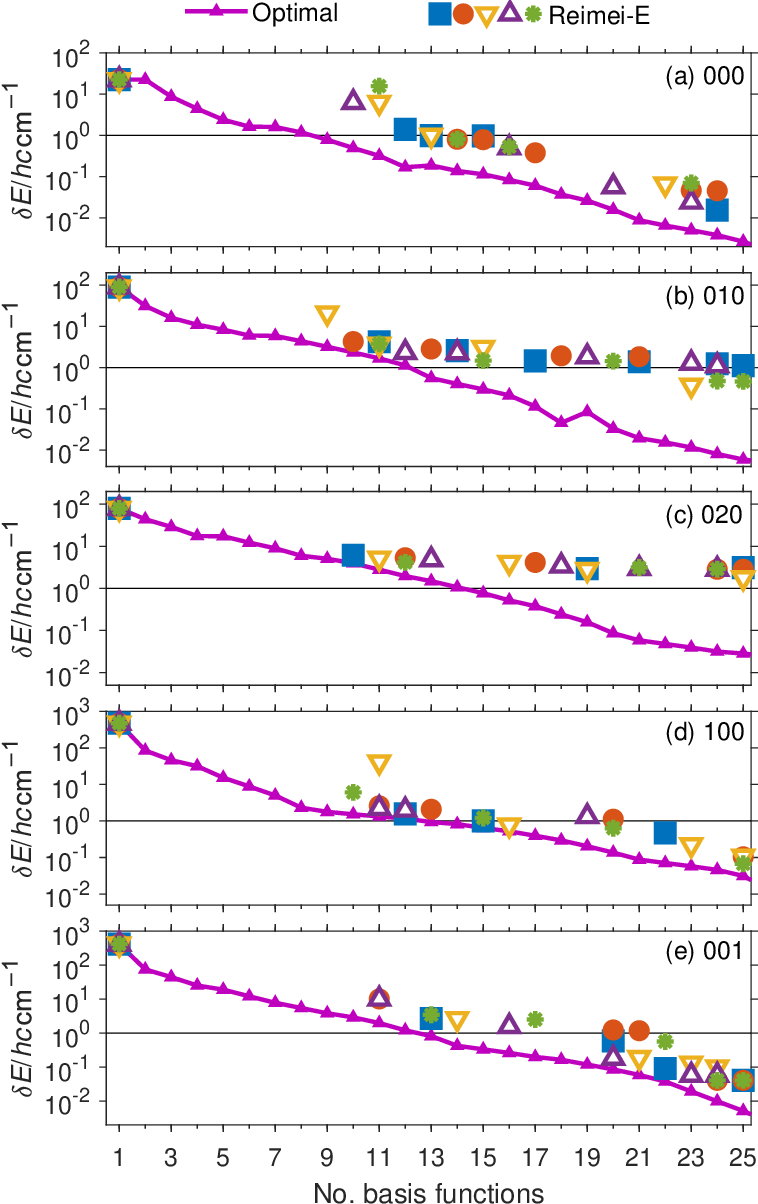}
\caption{\label{Fig2_Appendix}
Vibrational energy differences for $J=0$, $\vmax=3$, fixed $\tau=0.24$ fs, and $1\le \Ntrotter\le 6$. Each curve symbol ($\blacksquare$, {\Large $\bullet$}, $\triangledown$, $\vartriangle$, $\bm{\ast}$) represents a separate simulation by \Reimei{}-E, with $\Nshots=10^3$.
}
\end{figure}

%=========================================================================================================================
In Fig.~\ref{Fig2_Appendix}, we show vibrational energy differences $\delta E_{v_1v_2v_3}$ at $J=0$ and $\vmax=3$ 
(same conditions as in Fig.~\ref{Fig4}), obtained by sampling using \Reimei{}-E, a classical emulator of \Reimei{}. 
In the \Reimei{}-E emulator,  the simulations are performed on a classical computer using an accurate model of the noise in \Reimei{} \cite{RyanAnderson2021,QuantinuumNoiseModel} so that the final results are close to results obtained by a simulation using the real \Reimei{} quantum computer. 
We also show  the energy differences obtained using the optimal method. Each curve symbol in Fig.~\ref{Fig2_Appendix} represents a simulation using \Reimei{}-E and $\Nshots=10^3$. We can see that although there are differences in the simulations results due to the statistical nature of the sampling procedure, in all simulations we obtain energy differences of  a few \hcmminusone{} at about $|\Omega|=20$ basis functions.

%@@@@@@@@@@@@@@@@@@@@@@@@@@@@@@@@@@@@@@@@@@@@@@@@@@@@@@@@@@@@@@@@@@@@@@@@@@@@@@@@@@@@@@@
%@@@@@@@@@@@@@@@@@@@@@@@@@@@@@@@@@@@@@@@@@@@@@@@@@@@@@@@@@@@@@@@@@@@@@@@@@@@@@@@@@@@@@@@

\clearpage
%\newpage

%\bibliographystyle{apsrev41custom_italic}
%\bibliography{QuantumRefs,szt}%

%merlin.mbs apsrev4-1.bst 2010-07-25 4.21a (PWD, AO, DPC) hacked
%Control: key (0)
%Control: author (72) initials jnrlst
%Control: editor formatted (1) identically to author
%Control: production of article title (1) required
%Control: page (0) single
%Control: year (1) truncated
%Control: production of eprint (0) enabled
%

\end{document}